\documentclass[a4paper,11pt]{article}
\usepackage{cite}
\usepackage[top=1in,bottom=1in,left=1in,right=1in]{geometry}
\usepackage{amssymb,amsmath,amsthm}
\usepackage{graphicx}
\usepackage{appendix}
\usepackage{float}
\usepackage{xcolor}
\usepackage{hyperref}
\hypersetup{
  hidelinks,
  pdftitle={Realization of several related (2+1)-dimensional integrable systems via an anti-self-dual Yang--Mills reduction},
  pdfauthor={Shangshuai Li; Ken-ichi Maruno; Da-jun Zhang}
}

\numberwithin{equation}{section}
\DeclareMathOperator{\tr}{tr}
\DeclareMathOperator{\diag}{diag}
\DeclareMathOperator{\sech}{sech}

\theoremstyle{plain}
\newtheorem{theorem}{Theorem}
\newtheorem{lemma}{Lemma}
\newtheorem{proposition}{Proposition}

\theoremstyle{remark}
\newtheorem{remark}{Remark}

\usepackage{tikz}
\usetikzlibrary{positioning,shapes.misc}
\usepackage{graphicx}

\begin{document}

\title{\Large\bfseries
Realization of several related $(2+1)$-dimensional integrable systems via an anti-self-dual Yang--Mills reduction}

\author{Shangshuai Li$^{1,2}$\footnote{Email: lishangshuai@nbu.edu.cn}
\qquad Ken-ichi Maruno$^{3}$\footnote{Email: kmaruno@waseda.jp}
\qquad Da-jun Zhang$^{4,5}$\footnote{Corresponding author. Email: djzhang@staff.shu.edu.cn}
\\[5pt]
{\small $^1$School of Mathematics and Statistics, Ningbo University, Ningbo 315211, China} \\
{\small $^2$Graduate School of Mathematics, Nagoya University, Nagoya 464-8602, Japan} \\
{\small $^3$Department of Applied Mathematics, Faculty of Science and Engineering,} \\
{\small Waseda University, Tokyo 169-8555, Japan} \\
{\small $^4$Department of Mathematics, Shanghai University, Shanghai 200444, China} \\
{\small $^5$Newtouch Center for Mathematics of Shanghai University, Shanghai 200444, China}}

\date{}
\maketitle

\begin{abstract}
We derive several integrable equations in $(2+1)$ dimensions from the anti-self-dual Yang--Mills (ASDYM) equations with gauge group $G=\mathrm{GL}(2,\mathbb C)$ by imposing a single one-dimensional constraint on the ASDYM matrices. The resulting systems include the $(2+1)$-dimensional nonlinear Schr\"odinger (NLS) equation, a $(2+1)$-dimensional generalized derivative NLS family encompassing the Gerdjikov--Ivanov, Chen--Lee--Liu, and Kaup--Newell equations, and the $(2+1)$-dimensional Heisenberg ferromagnetic (HF) equation. The $(1+1)$-dimensional Wadati--Konno--Ichikawa equation is also recovered through a reciprocal transformation of the HF equation. The nonlinear variables and the associated gauge transformations are expressed explicitly in terms of the entries of the ASDYM matrices. The same one-dimensional constraint is implemented in the Cauchy matrix framework, yielding explicit solutions of the reduced integrable equations. In this formulation, the ASD Yang--Mills Lagrangian density is expressed in terms of a $\tau$-function; for suitable exponential data, this $\tau$-function becomes a Grammian-type $\tau$-function
of the Kadomtsev--Petviashvili equaton.
The studies are extended to $\mathrm{GL}(M,\mathbb C)$ at the matrix level.

\begin{description}
\item[Keywords:] Anti-self-dual Yang--Mills equation; integrable reduction; $(2+1)$-dimensional integrable system; Lagrangian density; $\tau$-function
\item[MSC2020:] 35Q51; 35Q55; 37K10; 37K40
\item[PACS:] 02.30.Ik; 02.30.Jr; 05.45.Yv
\end{description}
\end{abstract}


\section{Introduction}\label{sec-1}

	The anti-self-dual Yang--Mills (ASDYM) equations on four-dimensional complexified spacetime arise in both gauge theory and the theory of integrable systems \cite{Mason-Book}. On a Euclidean real slice, finite-action ASD fields saturate the Yang--Mills action bound in a localized domain and describe instantons \cite{Atiyah-1977,Atiyah-1978,Prasad-1980}. In complexified spacetime, the ASDYM equations provide a common source of many classical integrable systems \cite{Ward-1985}. Their connections with fiber bundles, soliton equations, and twistor geometry are reviewed in Parts I and II of \cite{Mason-Book}; see also \cite{Wu-1975}.

	Ward proposed that many lower-dimensional integrable systems can be obtained as reductions of the ASDYM equations \cite[p.~452]{Ward-1985}:
	\begin{quote}
		\textit{\ldots{} many (and perhaps all?) of the ordinary and partial differential equations that are regarded as being integrable or solvable
			may be obtained from the self-dual gauge field equations (or its generalizations) by reduction.}
	\end{quote}
	This statement is commonly referred to as the Ward conjecture. The Penrose--Ward correspondence identifies ASDYM gauge fields with holomorphic vector bundles over twistor space \cite{Ward-1977,Atiyah-1977,Atiyah-1979}. Symmetry reductions on twistor space yield lower-dimensional integrable equations, including the Korteweg--de Vries (KdV) and nonlinear Schr\"odinger (NLS) equations \cite{Mason-1989}. Alternatively, one may impose suitable ans\"atze on the gauge potentials so that the ASDYM Lax pair reduces to the corresponding Lax pairs of the reduced systems; this procedure also extends to integrable hierarchies (see \cite{Ablowitz-1993,Ablowitz-2003} and references therein).

	In previous work, we introduced a reduction framework based on the $J$- and $K$-matrix formulations of the ASDYM equations (see Section~\ref{sec-2-2}) and used it to derive the Fokas--Lenells (FL) equation \cite{LSS-FL-2025}. The reduction was shown to preserve solutions, thereby establishing a correspondence between ASDYM and FL solutions through the Cauchy matrix scheme \cite{LSS-2022,LSS-2023}. The same method was subsequently extended to $(1+1)$-dimensional integrable hierarchies, including the Ablowitz--Kaup--Newell--Segur (AKNS) hierarchy, the three standard derivative NLS (DNLS) hierarchies, and the Kadomtsev--Petviashvili (KP) and modified KP (mKP) hierarchies \cite{LMZ-2026}. In the present work, rather than generating $(1+1)$-dimensional hierarchies, we retain a $(2+1)$-dimensional reduction and show that several systems related by Miura, gauge, and reciprocal transformations, together with their Lax representations, Cauchy matrix solutions, and $\tau$-function structure, arise from the same one-dimensional symmetry.

We apply this reduction framework to several $(2+1)$-dimensional integrable equations. Some of these systems occur in hierarchies associated with toroidal algebras (cf.~\cite{Kakei-2002}). The equations considered below are related by Miura, gauge, and reciprocal transformations.
\begin{enumerate}
		\item \emph{The $(2+1)$-dimensional NLS equation.}
		\begin{align}\label{Zakharov-system}
			\mathrm{i}q_t=q_{xy}+2q\partial_x^{-1}(|q|^2)_{y},
		\end{align}
		where $\partial_x^{-1}f(x)=\int_{-\infty}^{x}f(x')\,\mathrm{d}x'$.
		This equation is a $(2+1)$-dimensional generalization of the standard $(1+1)$-dimensional NLS equation
		\begin{align}\label{NLS}
			\mathrm{i}q_t=q_{xx}+2|q|^2q.
		\end{align}
Note that two standard $(2+1)$-dimensional extensions of the $(1+1)$-dimensional NLS equation are the Zakharov equation above \cite{Zakharov-1980} and the Davey--Stewartson (DS) system \cite{Davey-1974}.
		\item \emph{The $(2+1)$-dimensional DNLS equations.}
One such equation is
		\begin{align}\label{Gerdjikov--Ivanov}
			\mathrm{i}q_t=q_{xy}-2\mathrm{i}q\partial_x^{-1}(qq_x^*)_y+2q\partial_x^{-1}(|q|^4)_y.
		\end{align}
		It is a $(2+1)$-dimensional generalization of the Gerdjikov--Ivanov (GI) equation \cite{Gerdjikov-1983}:
		\begin{align}\label{GI}
			\mathrm{i}q_t=q_{xx}-2\mathrm{i}q^2q_x^*+2|q|^4q,
		\end{align}
		which is one of the three standard DNLS equations.
		Applying the gauge transformation
		\begin{align}
			v=q\exp\left(-\mathrm{i}\gamma\int |q|^2\,\mathrm{d}x\right),\qquad \gamma\in\mathbb R,
		\end{align}
		yields the generalized DNLS (GDNLS) equation \cite{Ichikawa-1980,Kundu-1984,Kakei-1995}:
		\begin{align}\label{GDNLS}
			\mathrm{i}v_t=v_{xx}+2\mathrm{i}\gamma v_x|v|^2+2\mathrm{i}(\gamma-1)v^2v_x^*+(\gamma-1)(\gamma-2)|v|^4v.
		\end{align}
		For $\gamma=0,1,2$, this equation reduces to the GI, Chen--Lee--Liu (CLL) \cite{Chen-1979}, and Kaup--Newell (KN) \cite{Kaup-1978} equations, respectively. The same gauge transformation extends to $(2+1)$ dimensions and yields
		\begin{align}
			\mathrm{i}v_t=v_{xy}&+\mathrm{i}\gamma v\partial_x^{-1}(v_xv^*)_y+\mathrm{i}\gamma v_x\partial_x^{-1}(|v|^2)_y \notag \\
			 &+2\mathrm{i}(\gamma-1)v\partial_x^{-1}(vv_x^*)_y+(\gamma-1)(\gamma-2)v\partial_x^{-1}(|v|^4)_y.
		\end{align}
		To the best of our knowledge, this particular $(2+1)$-dimensional GDNLS equation does not appear to have been reported explicitly in this form. Other higher-dimensional DNLS-type systems and
hierarchies have been obtained by different constructions
\cite{Kakei-2002,Strachan-1992,Strachan-1992-2,Myrzakul-2000}.
We refer to this system as the $(2+1)$-dimensional GDNLS equation. The cases $\gamma=1$ and $\gamma=2$
give the $(2+1)$-dimensional CLL and KN equations, respectively:
			\begin{align}
				&\mathrm{i}v_t=v_{xy}+\mathrm{i} v\partial_x^{-1}(v_xv^*)_y+\mathrm{i}v_x\partial_x^{-1}(|v|^2)_y, \\
				&\mathrm{i}v_t=v_{xy}+2\mathrm{i}v(|v|^2)_y+2\mathrm{i}v_x\partial_x^{-1}(|v|^2)_y.
			\end{align}
		\item \emph{The $(2+1)$-dimensional Heisenberg ferromagnetic (HF) model.}
		\begin{subequations}\label{Myrzakulov-I}
			\begin{align}
				&\mathbf S_t=(\mathbf S\times\mathbf S_y+u\mathbf S)_x, \\
				&u_x=-\mathbf S\cdot(\mathbf S_x\times\mathbf S_y),
			\end{align}
		\end{subequations}
		where $\mathbf S:=(s_1,s_2,s_3)$ is a unit spin vector in $\mathbb R^3$. This system, known as the Myrzakulov-I equation \cite{Myrzakulov-1987}, is a $(2+1)$-dimensional generalization of the $(1+1)$-dimensional HF model:
		\begin{align}\label{LLE}
			\mathbf S_t=\mathbf S\times\mathbf S_{xx}.
		\end{align}
The HF model and the NLS equation are gauge equivalent in both $(1+1)$ and $(2+1)$ dimensions \cite{Myrzakulov-1987,Zakharov-1979}.
A reciprocal transformation \cite{Wadati-1982} gives the Wadati--Konno--Ichikawa (WKI) equation \cite{Wadati-1979-2,Shimizu-1980}:
		\begin{align}\label{Shimizu-Wadati}
			\hat q_T=\pm\mathrm{i}\left(\frac{\hat q}{\sqrt{1+|\hat q|^2}}\right)_{XX},
		\end{align}
		which belongs to the WKI hierarchy \cite{Wadati-1979}.
	\end{enumerate}

	ASDYM reductions of these systems have previously been studied separately. Reductions of the ASDYM equations to $(2+1)$-dimensional NLS and DNLS equations or hierarchies have been derived using twistor methods \cite{Strachan-1992,Strachan-1992-2} and bilinear identities \cite{Sasa-1998,Kakei-2002}, respectively. The $(1+1)$-dimensional NLS and DNLS hierarchies have also been derived directly from ASDYM matrix formulations (see Table~1 in Ref.~\cite{LMZ-2026}), and ASDYM reductions to Myrzakulov-type equations were studied in \cite{Myrzakul-2000}. In those constructions, the NLS, DNLS, and HF branches arise from different reductions or formalisms. Here, all branches considered below, together with their Miura, gauge, and reciprocal relations, follow from the same matrix constraint
(see equation \eqref{dimensional-reduction} in this paper).

	All the equations above are obtained from the ASDYM matrices by imposing the single one-dimensional constraint \eqref{dimensional-reduction}. The main results are as follows:
\begin{enumerate}
\item[(i)] A unified reduction of the $(2+1)$-dimensional NLS, DNLS, and HF equations from the $\mathrm{GL}(2,\mathbb C)$ ASDYM matrix formulations, including the GDNLS family displayed above.
\item[(ii)] Lax pairs for all the reduced systems, obtained from the ASDYM linear system under the same one-dimensional constraint (Section~\ref{sec-3-3}). In particular, this construction provides a zero-curvature representation for the $(2+1)$-dimensional GDNLS equation and establishes its integrability in the usual Lax-pair sense.
\item[(iii)] An expression for the reduced Yang--Mills Lagrangian density in terms of the $\tau$-function.
\end{enumerate}

The remainder of the paper is organized as follows. Section~\ref{sec-2} reviews the ASDYM equations and their matrix formulations. Section~\ref{sec-3} develops the reduction framework and derives the lower-dimensional integrable systems together with their Lax pairs. Section~\ref{sec-4} implements the same constraint in the Cauchy matrix framework and constructs explicit solutions of the reduced systems. Section~\ref{sec-5} summarizes the results and discusses directions for further work.
Note that the research of Section~\ref{sec-3} can be extended to
$\mathrm{GL}(M,\mathbb C)$ at the matrix level, which is presented in Appendix \ref{sec-3-4}.

\section{The ASDYM equations}\label{sec-2}

This section reviews the ASDYM equations using the conventions of Mason and Woodhouse \cite{Mason-Book}.
We work on a product of simply connected coordinate domains
and assume that the gauge potentials are holomorphic.

\subsection{Preliminaries on Yang--Mills theory}\label{sec-2-1}

Consider a 4-dimensional manifold $\mathbb C^4=(z^1,z^2,z^3,z^4)$ with metric matrix
	\begin{equation}\label{eta-mn}
		(\eta_{mn})_{4\times4}:=
		\begin{pmatrix}
			0 & 1 & 0 & 0 \\
			1 & 0 & 0 & 0 \\
			0 & 0 & 0 & -1 \\
			0 & 0 & -1 & 0
		\end{pmatrix},\qquad
		m,n=1,2,3,4.
	\end{equation}
	The corresponding metric is
	\begin{equation}\label{metric}
		 \mathrm{d}s^2=\eta_{mn}\mathrm{d}z^m\mathrm{d}z^n=2(\mathrm{d}z^1\mathrm{d}z^2-\mathrm{d}z^3\mathrm{d}z^4).
	\end{equation}

Let $G$ be a matrix Lie group with Lie algebra $\mathfrak g$, and let $\Phi$ be a vector in the fundamental representation of $G$. Under a gauge transformation,
	\begin{align}
		\Phi\mapsto \Phi'=g\Phi,\qquad g\in G.
	\end{align}
	Introduce $\mathfrak g$-valued gauge potentials $B_i$ and the covariant derivatives
	\begin{align}
		\nabla_i:=\partial_i+B_i,\qquad \partial_i:=\frac{\partial}{\partial z^i},\qquad B_i\in\mathfrak g.
	\end{align}
	These derivatives are covariant under the gauge transformation, i.e., $\nabla_i\Phi \mapsto \nabla'_i\Phi'=g(\nabla_i\Phi)$, where $\nabla'_i:=\partial_i+B'_i$ and $B'_i\in \mathfrak g$,
so the gauge potentials transform as
	\begin{align}\label{gauge-potential-transformation}
		B_i\mapsto B'_i=gB_ig^{-1}-(\partial_ig)g^{-1}.
	\end{align}
	The field strength is
	\begin{align}
		F_{ij}:=[\nabla_i,\nabla_j]=\partial_i B_j-\partial_j B_i+[B_i,B_j],
	\end{align}
	and transforms as
	\begin{align}
		F_{ij}\mapsto F'_{ij}:=[\nabla'_i,\nabla'_j]=gF_{ij}g^{-1}.
	\end{align}
The Yang--Mills Lagrangian density is
\begin{align}\label{Yang--Mills-Lagrangian}
\mathcal L=\tr(F_{ij}F^{ij}),
\end{align}
where $F^{ij}:=\eta^{ik}\eta^{jl}F_{kl}$ and $\eta^{mn}$ denotes the $(m,n)$ entry of the inverse of the metric matrix in \eqref{eta-mn}. For the metric in \eqref{eta-mn}, $\eta^{mn}=\eta_{mn}$.
By cyclicity of the trace,
\begin{align}\label{L-gauge-equivalent}
\mathcal L\mapsto \mathcal L'=\tr(gF_{ij}F^{ij}g^{-1})=\mathcal L,
\end{align}
so $\mathcal L$ is gauge invariant. This is the Lagrangian density of pure Yang--Mills theory.

\subsection{The ASD condition and its matrix formulations}\label{sec-2-2}

	The anti-self-dual (ASD) condition is given by
	\begin{equation}\label{ASD-condition}
		F_{ij}=-(*F)_{ij},\qquad (*F)_{ij}=\frac{1}{2}\epsilon_{ijkl}\eta^{ka}\eta^{lb}F_{ab},
	\end{equation}
	where $\epsilon_{ijkl}$ is the Levi--Civita tensor and the indices $i,j,k,l,a,b$ run over $\{1,2,3,4\}$. Expanding \eqref{ASD-condition} gives
	\begin{subequations}\label{ASD-condition-expand}
		\begin{align}
			\label{ASD-condition-expand-1}&F_{zw}=[\nabla_z,\nabla_w]=0,\qquad F_{\tilde z\tilde w}=[\nabla_{\tilde z},\nabla_{\tilde w}]=0, \\
			\label{ASD-condition-expand-2}&F_{z\tilde z}-F_{w\tilde w}=[\nabla_z,\nabla_{\tilde z}]-[\nabla_w,\nabla_{\tilde w}]=0.
		\end{align}
	\end{subequations}
	By the Frobenius theorem, the flatness conditions
	\eqref{ASD-condition-expand-1} ensure the local existence of
	invertible matrix-valued functions $h$ and $\tilde h$ satisfying
	\begin{equation}
		\nabla_z h=0,\qquad \nabla_w h=0,\qquad
		\nabla_{\tilde z}\tilde h=0,\qquad
		\nabla_{\tilde w}\tilde h=0.
	\end{equation}
	The gauge potentials are therefore
	\begin{subequations}\label{gauge-potential-h}
		\begin{align}
			&B_z=-(\partial_zh)h^{-1},\qquad B_w=-(\partial_wh)h^{-1}, \\
			&B_{\tilde z}=-(\partial_{\tilde z}\tilde h)\tilde h^{-1},\qquad B_{\tilde w}=-(\partial_{\tilde w}\tilde h)\tilde h^{-1}.
		\end{align}
	\end{subequations}
	Under \eqref{gauge-potential-transformation}, \eqref{gauge-potential-h} becomes
	\begin{subequations}\label{gauge-potential-transformed}
		\begin{align}
			&B'_z=-\left(\partial_z(gh)\right)(gh)^{-1},\qquad B'_w=-\left(\partial_w(gh)\right)(gh)^{-1}, \\
			&B'_{\tilde z}=-\left(\partial_{\tilde z}(g\tilde h)\right)(g\tilde h)^{-1},\qquad B'_{\tilde w}=-\left(\partial_{\tilde w}(g\tilde h)\right)(g\tilde h)^{-1}.
		\end{align}
	\end{subequations}
	Two convenient gauge choices are as follows:
	\begin{enumerate}
		\item \emph{The $J$-gauge and $K$-gauge: }

		Taking $g=\tilde h^{-1}$ in \eqref{gauge-potential-transformed}, defining $J:=\tilde h^{-1}h$, and writing the transformed potentials as $A_i$ gives the $J$-gauge
		\begin{subequations}\label{gauge-potential-J}
			\begin{align}
				&A_z=-(\partial_zJ)J^{-1}, \qquad A_{\tilde z}=0, \\
				&A_w=-(\partial_wJ)J^{-1},\qquad A_{\tilde w}=0.
			\end{align}
		\end{subequations}
		Substitution into the ASD condition gives the $J$-matrix formulation of the ASDYM equation
		\begin{equation}\label{J-matrix-formulation}
			\partial_{\tilde z}((\partial_zJ)J^{-1})-\partial_{\tilde w}((\partial_wJ)J^{-1})=0.
		\end{equation}
		Define the $K$-matrix by
		\begin{equation}\label{Miura-transformation-J-K}
			\partial_{\tilde z}K=-(\partial_wJ)J^{-1},\qquad \partial_{\tilde w}K=-(\partial_zJ)J^{-1},
		\end{equation}
		whose compatibility gives the $K$-matrix formulation
		\begin{equation}\label{K-matrix-formulation}
			\partial_z\partial_{\tilde z}K-\partial_w\partial_{\tilde w}K-[\partial_{\tilde z}K,\partial_{\tilde w}K]=0.
		\end{equation}

		\item \emph{The $P$-gauge and $Q$-gauge: }

		Taking $g=h^{-1}$ in \eqref{gauge-potential-transformed}, defining $P:=h^{-1}\tilde h=J^{-1}$, and writing the transformed potentials as $C_i$ gives the $P$-gauge
		\begin{subequations}\label{gauge-potential-C}
			\begin{align}
				&C_z=0,\qquad C_{\tilde z}=-(\partial_{\tilde z}P)P^{-1}=J^{-1}\partial_{\tilde z}J,
\label{gauge-potential-C-a}\\
				&C_w=0,\qquad C_{\tilde w}=-(\partial_{\tilde w}P)P^{-1}=J^{-1}\partial_{\tilde w}J.
\label{gauge-potential-C-b}
			\end{align}
		\end{subequations}
		Similarly, the $P$-matrix formulation is
		\begin{align}\label{P-matrix-formulation}
			\partial_{z}((\partial_{\tilde z}P)P^{-1})-\partial_{w}((\partial_{\tilde w}P)P^{-1})=0,
		\end{align}
		which is \eqref{J-matrix-formulation} with $J=P^{-1}$. Define the $Q$-matrix by
		\begin{equation}\label{Miura-transformation-P-Q}
			\partial_{w}Q=-(\partial_{\tilde z}P)P^{-1},\qquad \partial_{z}Q=-(\partial_{\tilde w}P)P^{-1},
		\end{equation}
		whose compatibility gives
		\begin{equation}
			\partial_z\partial_{\tilde z}Q-\partial_w\partial_{\tilde w}Q-[\partial_{z}Q,\partial_{w}Q]=0,
		\end{equation}
		which has the same form as the $K$-matrix formulation.
	\end{enumerate}

	The $J$- and $P$-matrix formulations take the form of the Yang equation \cite{Yang-1977,Brihaye-1978,Pohlmeyer-1980}. The $K$- and $Q$-matrix formulations have the same form and follow from the cubic Leznov--Mukhtarov--Parkes action \cite{Leznov-1987,Leznov-1987-2,Parkes-1992}. The resulting equation is also known as the Chalmers--Siegel equation \cite{Chalmers-1996}.

	These matrix formulations are also useful for evaluating the Lagrangian density \eqref{Yang--Mills-Lagrangian}. Under the ASD condition \eqref{ASD-condition-expand}, it becomes
	\begin{align}\label{Yang--Mills-Lagrangian-ASD}
		\mathcal L_{[\mathrm{ASD}]}=4\tr(F_{z\tilde w}F_{w\tilde z}-F_{z\tilde z}^2).
	\end{align}
	Using the $J$-gauge \eqref{gauge-potential-J} and $K$-gauge \eqref{Miura-transformation-J-K}, the relevant field strengths are
	\begin{subequations}
		\begin{align}
			&F_{z\tilde w}=-\partial_{\tilde w}A_z=\partial_{\tilde w}((\partial_zJ)J^{-1})=-\partial^2_{\tilde w}K, \\
			&F_{w\tilde z}=-\partial_{\tilde z}A_w=\partial_{\tilde z}((\partial_wJ)J^{-1})=-\partial^2_{\tilde z}K, \\
			&F_{z\tilde z}=-\partial_{\tilde z}A_z=\partial_{\tilde z}((\partial_zJ)J^{-1})=-\partial_{\tilde z}\partial_{\tilde w}K.
		\end{align}
	\end{subequations}
	Equation \eqref{Yang--Mills-Lagrangian-ASD} can therefore be written in terms of $J$ or $K$ as
	\begin{subequations}
		\begin{align}
			&\mathcal L_{[\mathrm{ASD}]}=4\tr\left(\partial_{\tilde z}((\partial_wJ)J^{-1})\partial_{\tilde w}((\partial_zJ)J^{-1})-\partial_{\tilde z}((\partial_zJ)J^{-1})\partial_{\tilde z}((\partial_zJ)J^{-1})\right)=:\mathcal L_{[J]}, \\
			\label{L-K}&\mathcal L_{[\mathrm{ASD}]}
=4\tr\left((\partial_{\tilde z}^2K)
(\partial_{\tilde w}^2K)-(\partial_{\tilde z\tilde w}K)(\partial_{\tilde z\tilde w}K)\right)=:\mathcal L_{[K]}.
		\end{align}
	\end{subequations}
	The $J$- and $K$-matrix formulations can thus be used to evaluate the ASD Yang--Mills Lagrangian density.
Section~\ref{sec-4-2} expresses this density directly in terms of a KP $\tau$-function.

\subsection{Gauge conditions for \texorpdfstring{$\mathrm{SU}(2)$}{SU(2)} in split signature}\label{sec-2-3}

We focus on the ASDYM equations in split signature $(+,+,-,-)$.
This real slice is commonly used in ASDYM reductions and
admits nonsingular soliton solutions for the $\mathrm{SU}(2)$ gauge group \cite{Hamanaka-2020,LSS-DL-2025,LSS-PRD-2025}.
These gauge conditions are used below for the NLS and HF reductions and are implemented in Section~\ref{sec-4-3}.

	We use the following coordinate transformation for the split-signature real slice (other choices are described in \cite{Mason-Book,Hamanaka-2020,LSS-DL-2025}):
	\begin{align}
		\begin{pmatrix}
			\tilde z & w \\
			\tilde w & z
		\end{pmatrix}=\frac{\sqrt2}{2}
		\begin{pmatrix}
			\mathrm{i}(x^2+x^0) & \mathrm{i}(x^1-x^3) \\
			\mathrm{i}(x^1+x^3) & \mathrm{i}(x^2-x^0)
		\end{pmatrix},\qquad (x^0,x^1,x^2,x^3)\in\mathbb R^4,
	\end{align}
	which leads to
	\begin{align}
		\mathrm{d}s^2=2(\mathrm{d}z\mathrm{d}\tilde z-\mathrm{d}w\mathrm{d}\tilde w)=(\mathrm{d}x^0)^2+(\mathrm{d}x^1)^2-(\mathrm{d}x^2)^2-(\mathrm{d}x^3)^2.
	\end{align}
	Thus $(z,\tilde z,w,\tilde w)\in\mathrm{i}\mathbb R^4$ are purely imaginary coordinates, and
	\begin{subequations}\label{new-gauge-potential}
		\begin{align}
			&A_z=-\frac{\sqrt2}{2}\mathrm{i}(A_2-A_0),\qquad A_{\tilde z}=-\frac{\sqrt2}{2}\mathrm{i}(A_2+A_0), \\
			&A_w=-\frac{\sqrt2}{2}\mathrm{i}(A_1-A_3),\qquad A_{\tilde w}=-\frac{\sqrt2}{2}\mathrm{i}(A_1+A_3).
		\end{align}
	\end{subequations}

	The corresponding special linear and unitary conditions are as follows:
	\begin{itemize}
		\item \textbf{Special linear condition.} The gauge potentials satisfy
		\begin{align}
			\tr(A_i)=0,\qquad i\in\{0,1,2,3\}.
		\end{align}
		Together with \eqref{new-gauge-potential}, this gives
		\begin{align}
			\tr(A_\alpha)=0,\qquad \alpha\in\{z,\tilde z,w,\tilde w\}.
		\end{align}
		Hence
		\begin{subequations}
			\begin{align}
				&\tr(A_z)=\tr(\partial_{\tilde w}K)=-\tr((\partial_zJ)J^{-1})
=-\partial_z(\ln\det(J))=0, \\
				&\tr(A_w)=\tr(\partial_{\tilde z}K)=-\tr((\partial_wJ)J^{-1})
=-\partial_w(\ln\det(J))=0.
			\end{align}
		\end{subequations}
		Consequently, $\det(J)$ is independent of $z$ and $w$, while $\tr(K)$ is
independent of $\tilde z$ and $\tilde w$. The remaining dependence can be absorbed by the residual
freedom in $h,\tilde h$ and in the integration ``constants'' of \eqref{Miura-transformation-J-K}.
Without loss of generality, we therefore assume that $\tr(K)$ and
$\det(J)$ are constants. As emphasized in Remark~\ref{rem-SL2-SU2}(i) below, constancy
by itself only makes the gauge potentials $\mathfrak{sl}(2,\mathbb C)$-valued.
		\item \textbf{Unitary condition.} The gauge potentials satisfy
		\begin{align}
			A_i=-A_i^\dagger,\qquad i\in\{0,1,2,3\}.
		\end{align}
		Together with \eqref{new-gauge-potential}, this gives
		\begin{align}
			A_\alpha=A_\alpha^\dagger,\qquad \alpha\in\{z,\tilde z,w,\tilde w\}.
		\end{align}
		Hence the gauge conditions are
		\begin{subequations}
			\begin{align}
				&\partial_\alpha K=(\partial_{\alpha} K)^\dagger=-\partial_{\alpha}K^\dagger\qquad \Rightarrow\qquad K=-K^\dagger, \\
				&(\partial_\beta J)J^{-1}=((\partial_\beta J)J^{-1})^\dagger=-J^{-\dagger}(\partial_\beta J^\dagger)\qquad \Rightarrow\qquad J=J^{-\dagger},
			\end{align}
		\end{subequations}
		where $\alpha\in\{\tilde z,\tilde w\}$, $\beta\in\{z,w\}$ and $J^{-\dagger}$ denotes $(J^{-1})^\dagger$.
	\end{itemize}

We distinguish the following three levels, which will be used throughout the paper.

\begin{remark}\label{rem-SL2-SU2}
\emph{(i) Lie-algebra level.}
If $\det(J)$ and $\tr(K)$ are constants, then
$\tr(A_z)=\tr(\partial_{\tilde w}K)=-\partial_z\ln\det(J)=0$ and likewise
$\tr(A_w)=0$, while $A_{\tilde z}=A_{\tilde w}=0$ in the $J$-gauge. Hence all the gauge
potentials are traceless,
	\begin{align}
		A_\alpha\in\mathfrak{sl}(2,\mathbb C),\qquad \alpha\in\{z,\tilde z,w,\tilde w\},
	\end{align}
i.e., the connection is $\mathfrak{sl}(2,\mathbb C)$-valued.

\emph{(ii) Group level.}
Since $\det(J)$, $\det(P)$, $\tr(K)$ and $\tr(Q)$ are
constants, one may normalize
	\begin{align}\label{normalization-SU2}
		J\mapsto\frac{J}{\sqrt{\det(J)}},\qquad
		P\mapsto\frac{P}{\sqrt{\det(P)}},\qquad
		K\mapsto K-\frac{1}{2}\tr(K)I,\qquad
		Q\mapsto Q-\frac{1}{2}\tr(Q)I,
	\end{align}
where the normalized matrices still
satisfy $P=J^{-1}$.
This normalization affects neither the gauge potentials $A_\alpha$ and $C_\alpha$, nor the matrix $S$
introduced in \eqref{definition-S}, nor any of the reduced equations derived below, since these
depend on $J,P$ only through $(\partial_\alpha J)J^{-1}$, $J^{-1}\sigma_3J$ and on $K,Q$ only
through their derivatives. After \eqref{normalization-SU2} one has
	\begin{align}
		\det(J)=\det(P)=1,\qquad \tr(K)=\tr(Q)=0,
	\end{align}
	i.e., $J,P\in\mathrm{SL}(2,\mathbb C)$ and $K,Q\in\mathfrak{sl}(2,\mathbb C)$.

\emph{(iii) Reality level.}
Imposing in addition the unitary conditions $J^\dagger J=I$ and $K^\dagger=-K$ obtained above
restricts $J$ to the intersection $\mathrm{SL}(2,\mathbb C)\cap\mathrm{U}(2)$, that is, $J\in\mathrm{SU}(2)$.

\end{remark}

Accordingly, throughout the paper, the phrase ``the $\mathrm{SL}(2)$ (respectively, $\mathrm{SU}(2)$) gauge condition'' means that step (ii) (respectively, steps (ii) and (iii)) of Remark~\ref{rem-SL2-SU2} has been carried out.

\section{The ASDYM reduction framework}\label{sec-3}

We now impose a dimensional reduction under which the four-dimensional ASDYM equations reduce to $(2+1)$-dimensional integrable systems.

Throughout the paper, we take $G=\mathrm{GL}(2,\mathbb C)$; hence $J$, $K$, $P$, and $Q$ are $2\times2$ matrices:
\begin{align}\label{J-K-P-Q-expand}
	J=
	\begin{pmatrix}
		J_{11} & J_{12} \\
		J_{21} & J_{22}
	\end{pmatrix},\qquad
	K=
	\begin{pmatrix}
		K_{11} & K_{12} \\
		K_{21} & K_{22}
	\end{pmatrix},\qquad
	P=
	\begin{pmatrix}
		P_{11} & P_{12} \\
		P_{21} & P_{22}
	\end{pmatrix},\qquad
	Q=
	\begin{pmatrix}
		Q_{11} & Q_{12} \\
		Q_{21} & Q_{22}
	\end{pmatrix}.
\end{align}
We impose the following constraint on the $\tilde w$-dependence of these matrices (cf.~Eqs.~(2.5) and (2.7) of Ref.~\cite{LSS-FL-2025}):
\begin{align}\label{dimensional-reduction}
	\partial_{\tilde w}J=[J,C],\qquad
	\partial_{\tilde w}P=[P,C],\qquad
	\partial_{\tilde w}K=[K,C],\qquad
	\partial_{\tilde w}Q=[Q,C],
\end{align}
where $C=\sigma_3/2$ and $\sigma_3:=\diag(1,-1)$ is the third Pauli matrix.
This is the reduction constraint used throughout this section.

\begin{remark}
	The constraint \eqref{dimensional-reduction} has the following geometric interpretation. Since $C$ is constant,
	each equation integrates to
\begin{align}\label{dimensional-reduction-integrated}
	X(z,\tilde z,w,\tilde w)=e^{-\tilde wC}\,\widehat X(z,\tilde z,w)\,e^{\tilde wC},\qquad
		X\in\{J,K,P,Q\},\qquad \partial_{\tilde w}\widehat X=0.
\end{align}
In other words, the ASDYM matrices are invariant under the translation generated by the null
	Killing vector $\partial_{\tilde w}$ up to the one-parameter gauge transformation
		$g(\tilde w)=e^{-\tilde wC}$, whose infinitesimal generator $C\in\mathfrak{sl}(2,\mathbb C)$ plays
	the role of the Higgs field associated with this symmetry reduction
	(cf.~the general reduction scheme in Chapter 4 of \cite{Mason-Book}).
	Thus \eqref{dimensional-reduction} is a symmetry reduction along $\partial_{\tilde w}$ with a
	nontrivial Higgs field.

\end{remark}

\subsection{\texorpdfstring{Reduction to the $(2+1)$-dimensional NLS and $(2+1)$-dimensional DNLS equations}{Reduction to the (2+1)-dimensional NLS and (2+1)-dimensional DNLS equations}}\label{sec-3-1}

First, the constraint $\partial_{\tilde w}K=[K,C]$ in \eqref{dimensional-reduction}
reduces the $K$-matrix formulation \eqref{K-matrix-formulation} to the following 3-dimensional system:
	\begin{subequations}
		\begin{align}
			&\partial_zK_{11}=K_{12}K_{21},\qquad \partial_zK_{22}=-K_{21}K_{12}, \label{3.3a}\\
			&\partial_z\partial_{\tilde z}K_{12}=-\partial_{w}K_{12}-(\partial_{\tilde z}K_{11})K_{12}+K_{12}(\partial_{\tilde z}K_{22}), \label{3.3b}\\
			&\partial_z\partial_{\tilde z}K_{21}=\partial_{w}K_{21}
+(\partial_{\tilde z}K_{22})K_{21}-K_{21}(\partial_{\tilde z}K_{11}). \label{3.3c}
		\end{align}
	\end{subequations}
Set
	\begin{align}
		(r,q):=(K_{21},-K_{12}).
	\end{align}
Then \eqref{3.3a} gives $K_{11}=-K_{22}=-\partial_{z}^{-1}(qr)$.\footnote{Here we set the integration constants to be zero.}
Substitution into \eqref{3.3b} and \eqref{3.3c} yields the following $(2+1)$-dimensional NLS system:
	\begin{subequations}\label{NLS-system}
		\begin{align}
			\label{NLS-system-1}&r_w=r_{z\tilde z}-2r\partial_z^{-1}(qr)_{\tilde z},	 \\
			\label{NLS-system-2}&q_w=-q_{z\tilde z}+2q\partial_z^{-1}(rq)_{\tilde z}.
		\end{align}
	\end{subequations}
	Under the coordinate transformation $(x,y,t):=(\mathrm{i}z,\mathrm{i}\tilde z,\mathrm{i}w)\in\mathbb R^3$ and the conjugate reduction $q=r^*$,
the system reduces to the Zakharov equation \eqref{Zakharov-system}.\footnote{Throughout this section, this transformation maps the complex ASDYM coordinates to real spacetime.}

Next, consider the second equation in \eqref{Miura-transformation-J-K} together with
$\partial_{\tilde w}K=[K,C]$.
It follows that $\partial_zJ=-[K,C]J$. Expanding this matrix equation gives (cf.~Eq.~(2.10a) of Ref.~\cite{LSS-FL-2025})
	\begin{align}\label{3.6}
		\partial_zJ_{11}=K_{12}J_{21},\qquad \partial_zJ_{12}=K_{12}J_{22},\qquad
\partial_zJ_{21}=-K_{21}J_{11},\qquad \partial_zJ_{22}=-K_{21}J_{12}.
	\end{align}
	Set
	\begin{align}\label{p-def}
		p:=J_{21}J_{11}^{-1}.
	\end{align}
	Differentiating $p$ with respect to $z$ gives
	\begin{align}\label{Miura-NLS-GI}
		p_z=\frac{\partial_zJ_{21}}{J_{11}}-\frac{J_{21}\partial_zJ_{11}}{J_{11}^2}
=-\frac{K_{21}J_{11}}{J_{11}}-p\frac{K_{12}J_{21}}{J_{11}}=-r+p^2q.
	\end{align}
This recovers the standard Miura transformation between the $(2+1)$-dimensional NLS and GI systems (cf.~Eq.~(2.28) of Ref.~\cite{Kakei-2004} for the $(1+1)$-dimensional case). The corresponding $(2+1)$-dimensional GI system is
	\begin{subequations}\label{GI-system}
		\begin{align}
			&p_w=p_{z\tilde z}-2p\partial_z^{-1}(pq_z+p^2q^2)_{\tilde z}, \label{GI-system-a} \\
			&q_w=-q_{z\tilde z}-2q\partial_z^{-1}(p_zq-p^2q^2)_{\tilde z}. \label{GI-system-b}
		\end{align}
	\end{subequations}

\begin{remark}
	Substituting \( r = -p_z + p^2q \) from \eqref{Miura-NLS-GI} into \eqref{NLS-system-2} gives \eqref{GI-system-b} directly.
	Equation \eqref{GI-system-a} can be obtained just  directly, without inverting any
	differential operator, as follows.
	The first equation in \eqref{Miura-transformation-J-K}
	gives $\partial_wJ=-(\partial_{\tilde z}K)J$. Its $(1,1)$ and $(2,1)$ entries are
	\[
	\partial_wJ_{11}=-(\partial_{\tilde z}K_{11})J_{11}-(\partial_{\tilde z}K_{12})J_{21},\qquad
	\partial_wJ_{21}=-(\partial_{\tilde z}K_{21})J_{11}-(\partial_{\tilde z}K_{22})J_{21}.
	\]
	Hence, with
\begin{equation}
p=J_{21}J_{11}^{-1},~~~ (r,q)=(K_{21},-K_{12})
\end{equation}
 and
	$K_{11}=-K_{22}=-\partial_z^{-1}(qr)$,
	\[
	p_w=\frac{\partial_wJ_{21}}{J_{11}}-p\frac{\partial_wJ_{11}}{J_{11}}
	=-r_{\tilde z}-2p\,\partial_z^{-1}(qr)_{\tilde z}-p^2q_{\tilde z}.
	\]
	Substituting $r=-p_z+p^2q$ and $qr=-p_zq+p^2q^2$ and using the elementary identity
	$\partial_z^{-1}(pq_z)_{\tilde z}+\partial_z^{-1}(p_zq)_{\tilde z}=(pq)_{\tilde z}$
	yields \eqref{GI-system-a}.
\end{remark}

Under the coordinate transformation $(x,y,t):=(\mathrm{i}z,\mathrm{i}\tilde z,\mathrm{i}w)\in\mathbb R^3$ and the conjugate reduction $q=p^*$, the system reduces to the $(2+1)$-dimensional GI equation \eqref{Gerdjikov--Ivanov}.

Next, introduce the gauge factor (cf.~Eq.~(3.21) of Ref.~\cite{LMZ-2026})
	\begin{align}\label{gauge-trans-1}
		s:=\exp(-\partial_z^{-1}(pq))=J_{11}.
	\end{align}
The second equality follows from \eqref{3.6}, which gives $\partial_z\ln J_{11}=K_{12}J_{21}/J_{11}=-pq$, after choosing the integration constant in $\partial_z^{-1}$ so that $s=J_{11}$.
Define $(u,v)$ by the gauge transformation
	\begin{align}\label{gauge-trans-2}
		(u,v):=(ps^\gamma,s^{-\gamma}q)=(J_{21}J_{11}^{\gamma-1},-J_{11}^{-\gamma}K_{12}),\qquad \gamma\in\mathbb C.
	\end{align}
A direct calculation yields the following $(2+1)$-dimensional GDNLS system:
	\begin{subequations}\label{GDNLS-system}
		\begin{align}
			&u_w=u_{z\tilde z}+\gamma u\partial_z^{-1}(u_zv)_{\tilde z}+\gamma u_z\partial_z^{-1}(uv)_{\tilde z}+2(\gamma-1)u\partial_z^{-1}(uv_z)_{\tilde z}-(\gamma-1)(\gamma-2)u\partial_z^{-1}(u^2v^2)_{\tilde z}, \\
			&v_w=-v_{z\tilde z}+\gamma v\partial_z^{-1}(uv_z)_{\tilde z}+\gamma v_z\partial_z^{-1}(uv)_{\tilde z}+2(\gamma-1)v\partial_z^{-1}(u_zv)_{\tilde z}+(\gamma-1)(\gamma-2)v\partial_z^{-1}(u^2v^2)_{\tilde z},
		\end{align}
	\end{subequations}
provided that the pair $(p,q)$ satisfies the $(2+1)$-dimensional GI system \eqref{GI-system}.

	For special values of $\gamma$, the GDNLS system \eqref{GDNLS-system} reduces to the standard DNLS systems.
For example, $(u|_{\gamma=0},v|_{\gamma=0})=(p,q)$ yields the GI system \eqref{GI-system}.
In particular:
	\begin{itemize}
		\item When $\gamma=1$, $(u|_{\gamma=1},v|_{\gamma=1})=(ps,s^{-1}q)=(J_{21},-J_{11}^{-1}K_{12})$ satisfies the $(2+1)$-dimensional CLL system
		\begin{subequations}
			\begin{align}
				&u_w=u_{z\tilde z}+u\partial_z^{-1}(u_zv)_{\tilde z}+u_z\partial_z^{-1}(uv)_{\tilde z}, \\
				&v_w=-v_{z\tilde z}+ v\partial_z^{-1}(uv_z)_{\tilde z}+ v_z\partial_z^{-1}(uv)_{\tilde z}.
			\end{align}
		\end{subequations}
		\item When $\gamma=2$, $(u|_{\gamma=2},v|_{\gamma=2})=(ps^2,s^{-2}q)=(J_{21}J_{11},-J_{11}^{-2}K_{12})$ satisfies the $(2+1)$-dimensional KN system
		\begin{subequations}
			\begin{align}
				&u_w=u_{z\tilde z}+2(u\partial_z^{-1}(uv)_{\tilde z})_z, \\
				&v_w=-v_{z\tilde z}+2(v\partial_z^{-1}(uv)_{\tilde z})_z.
			\end{align}
		\end{subequations}
	\end{itemize}

Under the coordinate transformation $(x,y,t):=(\mathrm{i}z,\mathrm{i}\tilde z,\mathrm{i}w)\in\mathbb R^3$, the conjugate reduction $u=v^*$ (equivalently, $q=p^*$), and the restrictions $\gamma\in\mathbb R$ and $|s|^2=1$, the GDNLS system \eqref{GDNLS-system} becomes
	\begin{align}\label{GDNLS-2p1}
		\mathrm{i}v_t=v_{xy}&+\mathrm{i}\gamma v\partial_x^{-1}(v_xv^*)_y+\mathrm{i}\gamma v_x\partial_x^{-1}(|v|^2)_y \notag \\
		&+2\mathrm{i}(\gamma-1)v\partial_x^{-1}(vv_x^*)_y+(\gamma-1)(\gamma-2)v\partial_x^{-1}(|v|^4)_y.
	\end{align}

\begin{remark}\label{rem-2}
		Suppose that the normalization \eqref{normalization-SU2} has been carried out, so that
	$\det(J)=1$ and $\tr(K)=\tr(Q)=0$ (Remark~\ref{rem-SL2-SU2}(ii)).
The second relation in \eqref{Miura-transformation-P-Q}, together with \eqref{gauge-potential-C-b}, gives
\begin{equation}\label{3.16}
\partial_zQ=-(\partial_{\tilde w}P)P^{-1}=J^{-1}(\partial_{\tilde w}J).
\end{equation}
Imposing the first constraint equation in \eqref{dimensional-reduction} yields
$\partial_zQ=J^{-1}[J,C]$,
whose four entries are (cf.~Eq.~(2.10b) of Ref.~\cite{LSS-FL-2025})
\begin{align}
\partial_zQ_{11}=-J_{12}J_{21},\qquad \partial_zQ_{12}=-J_{22}J_{12},\qquad
\partial_zQ_{21}=J_{11}J_{21},\qquad \partial_zQ_{22}=J_{21}J_{12}.
\end{align}
		Hence
		\begin{align}
			\partial_zQ_{21}=J_{11}J_{21}=u|_{\gamma=2},\qquad
\partial_z(J_{11}^{-1}J_{12})=J_{11}^{-2}K_{12}=-v|_{\gamma=2},
		\end{align}
where the second equation is a consequence of \eqref{3.6}.
		Thus,
\begin{equation}
(\mu,\nu)=(Q_{21},-J_{11}^{-1}J_{12})=(\partial_z^{-1}(u|_{\gamma=2}),\partial_z^{-1}(v|_{\gamma=2})),
\end{equation}
are potential variables for the KN system.
These relations recover the FL variables used in the ASDYM reduction of \cite{LSS-FL-2025}.
	\end{remark}

\subsection{\texorpdfstring{Reduction to the $(2+1)$-dimensional HF equation and the WKI equation}{Reduction to the (2+1)-dimensional HF equation and the WKI equation}}\label{sec-3-2}

Define the matrix $S$ by
	\begin{align}\label{definition-S}
		S:=J^{-1}\sigma_3J=
		\begin{pmatrix}
			f & g \\
			e & -f
		\end{pmatrix},
	\end{align}
	where $f^2+eg=1$, with components
	\begin{align}
		e=-\frac{2J_{11}J_{21}}{\det(J)},\qquad
		f=\frac{\det(J)+2J_{12}J_{21}}{\det(J)}
		=\frac{J_{11}J_{22}+J_{12}J_{21}}{\det(J)},\qquad
		g=\frac{2J_{12}J_{22}}{\det(J)}.
	\end{align}
	This matrix satisfies $S^2=I$, $\det(S)=-1$, and $\tr(S)=0$.

Using the constraint \eqref{dimensional-reduction}, we write $\partial_zQ$ in terms of $S$ as
	\begin{align}\label{Q-S}
		\partial_zQ=J^{-1}[J,C]=C-J^{-1}CJ=C-\frac{1}{2}S.
	\end{align}
Similarly, the first relation in \eqref{Miura-transformation-P-Q}, together with \eqref{gauge-potential-C-a}, gives
\begin{equation}\label{Q-w}
\partial_{w}Q=-(\partial_{\tilde z}P)P^{-1}=J^{-1}(\partial_{\tilde z}J).
\end{equation}
Differentiating $S$ in \eqref{definition-S} with respect to $\tilde z$ gives
\begin{align}\label{S-zt-evo}
\partial_{\tilde z}S=-J^{-1}(\partial_{\tilde z}J)J^{-1}\sigma_3 J+J^{-1}\sigma_3(\partial_{\tilde z}J)
=[J^{-1}\sigma_3 J, J^{-1}(\partial_{\tilde z}J)]
=[S,\partial_wQ].
\end{align}

We next derive the $(2+1)$-dimensional HF equation using the following standard decomposition of $\mathfrak{sl}(2,\mathbb C)$-valued matrices \cite[Chapter~V]{Helgason-2001}.

\begin{lemma}\label{lem-1}
		Let $S$ be a traceless matrix satisfying $S^2=I$. Every $A\in\mathfrak{sl}(2,\mathbb C)$
admits the unique decomposition
		\begin{align}\label{3.25}
			A=A_{\perp}+A_{\parallel},\qquad A_{\perp}=\frac{1}{4}[S,[S,A]],\qquad
A_{\parallel}=\frac{1}{2}\tr(AS)S,
		\end{align}
		where $A_\perp$ anticommutes with $S$ ($SA_\perp=-A_\perp S$), whereas $A_\parallel$ commutes with $S$ ($SA_\parallel=A_\parallel S$).
	\end{lemma}

The reduction is summarized in the following theorem.

\begin{theorem}\label{thm-1}
Let $S$ be defined by \eqref{definition-S}, with the ASDYM matrices satisfying the constraint
\eqref{dimensional-reduction}. Then $S$ satisfies the $(2+1)$-dimensional HF system
		\begin{subequations}
			\begin{align}
				 \label{Heisenberg-1}&\partial_wS+2\kappa(\partial_zS)+\frac{1}{2}[S,\partial_z\partial_{\tilde z}S]=0, \\
				\label{Heisenberg-2}&\partial_z\kappa=-\frac{1}{8}\tr([\partial_zS,\partial_{\tilde z}S]S),
			\end{align}
		\end{subequations}
where $\kappa$ is defined by \eqref{3.27}. With respect to the decomposition in Lemma~\ref{lem-1}, equations \eqref{Heisenberg-1} and \eqref{Heisenberg-2} are
$(\partial_w S)_{\perp}$ and $(\partial_w S)_{\parallel}$
of the evolution equation for $S$ with respect to $w$, respectively.
\end{theorem}

\begin{proof}
We use the decomposition in Lemma~\ref{lem-1}.
First, decompose $\partial_wQ$ with respect to $S$:
\begin{align}\label{3.27}
			\partial_wQ=\frac{1}{4}[S,\partial_{\tilde z}S]+\kappa S,\qquad \kappa:=\frac{1}{2}\tr((\partial_wQ)S).
\end{align}
Using \eqref{Q-S} and \eqref{3.27}, the compatibility condition $\partial_z(\partial_wQ)=\partial_w(\partial_zQ)$ gives
\begin{align}
-\partial_wS=	2\partial_w(\partial_zQ) &	= 2\partial_z(\partial_wQ) \nonumber \\
&= \frac{1}{2}[\partial_zS,\partial_{\tilde z}S]
+\frac{1}{2}[S,\partial_z\partial_{\tilde z}S]+2(\partial_z\kappa)S+2\kappa(\partial_zS). \label{3.28}
\end{align}

Decompose $B:=-\partial_wS\in\mathfrak{sl}(2,\mathbb C)$ as $B=B_{\perp}+B_{\parallel}$ using Lemma~\ref{lem-1}.
Differentiating $S^2=I$ yields $(\partial_\alpha S)S+S(\partial_{\alpha}S)=0$, so
$\partial_\alpha S$ anticommutes with $S$
for any coordinate derivative $\partial_{\alpha}$, where $\alpha\in \{z, \tilde z, w, \tilde w\}$.
Hence
\begin{equation}\label{3.29}
B_{\perp}=-\partial_wS,\qquad B_{\parallel}=0.
\end{equation}
Using \eqref{3.25} and \eqref{3.28}, we obtain
\begin{align*}
B_{\perp} &=\frac{1}{4}[S,[S, -\partial_wS]]\\
& =\frac{1}{2}[S,\partial_z\partial_{\tilde z}S]+2\kappa(\partial_zS),
\end{align*}
where we used \eqref{3.28} and the anticommutation relation
$(\partial_\alpha S)S=-S(\partial_{\alpha}S)$.
Therefore,
\begin{align}\label{3.30}
-\partial_wS=\frac{1}{2}[S,\partial_z\partial_{\tilde z}S]+2\kappa(\partial_zS),
\end{align}
which is \eqref{Heisenberg-1}.
Comparing \eqref{3.28} and \eqref{3.30} gives
		\begin{align*}
			B_{\parallel}=\frac{1}{2}[\partial_zS,\partial_{\tilde z}S]+2(\partial_z\kappa)S=0.
		\end{align*}
Multiplying $B_{\parallel}=0$ on the right by $S$ and using $S^2=I$ gives
		\begin{align*}
			(\partial_z\kappa)I=-\frac{1}{4}[\partial_zS,\partial_{\tilde z}S]S.
		\end{align*}
Taking the trace and using $\tr(I)=2$ then yields
		\begin{align*}
			\partial_z\kappa=-\frac{1}{8}\tr([\partial_zS,\partial_{\tilde z}S]S),
		\end{align*}
		which is \eqref{Heisenberg-2}.
This proves the theorem.

\end{proof}

Under the coordinate transformation $(x,y,t):=(\mathrm{i}z,\mathrm{i}\tilde z,\mathrm{i}w)\in\mathbb R^3$, the system becomes
	\begin{subequations}\label{Heisenberg-model}
		\begin{align}
			&\mathrm{i}S_t-\mathrm{i}u S_x-\frac{1}{2}[S,S_{xy}]=0, \label{Heisenberg-model-1} \\
			&\mathrm{i}u_x=-\frac{1}{4}\tr([S_x,S_y]S), \label{Heisenberg-model-2}
		\end{align}
	\end{subequations}
	where $u:=-2\kappa=-\tr((\partial_wQ)S)$.

	Impose the Hermitian condition $S=S^\dagger$ and write $\mathbf S:=(s_1,s_2,s_3)$ with
$s_1^2+s_2^2+s_3^2=1$. Then $S$ is expressed in terms of the Pauli matrices as
	\begin{align}
S=s_1\sigma_1+s_2\sigma_2+s_3\sigma_3,
	\end{align}
where $\sigma_1,\sigma_2,\sigma_3$ are Pauli matrices,
\[\sigma_1=\begin{pmatrix}
			0 & 1 \\
			1 & 0
		\end{pmatrix},\qquad
\sigma_2=\begin{pmatrix}
			0 & -\mathrm{i} \\
			\mathrm{i} & 0
		\end{pmatrix},\qquad
\sigma_3=\begin{pmatrix}
			1 & 0 \\
			0 & -1
		\end{pmatrix}.
\]
	This gives
	\begin{align}
		S=
		\begin{pmatrix}
			f & g \\
			g^* & -f
		\end{pmatrix}
		=
		\begin{pmatrix}
			s_3 & s_1-\mathrm{i}s_2 \\
			s_1+\mathrm{i}s_2 & -s_3
		\end{pmatrix}.
	\end{align}
	Equation \eqref{Heisenberg-model} is then equivalent to the Myrzakulov-I equation \eqref{Myrzakulov-I} for \(\mathbf S\).

		Consider the $(1+1)$-dimensional reduction of \eqref{Heisenberg-model} obtained by setting $\partial_x=\partial_y$ and $u=0$.
		This yields
		\begin{align}\label{HFM-1p1}
			\mathrm{i}S_t=\frac{1}{2}[S,S_{xx}].
		\end{align}
		\begin{remark}\label{rem-spatial-translation}
			After applying the symmetry $\partial_x=\partial_y$ to \eqref{Heisenberg-model}, equation \eqref{Heisenberg-model-2} gives $u_x=0$; hence, in general, $u=u(t)$. In this case, equation \eqref{Heisenberg-model-1} reduces to
			\[
			\mathrm{i}(\partial_t-u(t)\partial_x)S-\frac{1}{2}[S,S_{xx}]=0.
			\]
			This advection term is removed by the time-dependent spatial translation
			$x'=x+a(t)$, $t'=t$, where $a'(t)=u(t)$. Indeed, if
			$\widetilde S(x',t'):=S(x'-a(t'),t')$, then
			$\widetilde S_{t'}=S_t-u(t')S_x$ and $\widetilde S_{x'x'}=S_{xx}$, so the equation becomes
			\[
			\mathrm{i}\widetilde S_{t'}-\frac{1}{2}[\widetilde S,\widetilde S_{x'x'}]=0.
			\]
			Thus one may work in the translated coordinate and set $u=0$ without loss of generality for this reduction.
		\end{remark}

		The $(1,1)$ entry of \eqref{HFM-1p1} gives the conservation law
		\begin{align}\label{f-conservation}
			f_t+\partial_x\mathcal H=0,\qquad
			\mathcal H:=\frac{\mathrm{i}}{2}(g e_x-g_xe).
		\end{align}
		Hence, on a simply connected domain on which $f\neq0$, the reciprocal transformation is
		defined by the closed one-form (cf.~Eqs.~(4.8) and (4.9) of Ref.~\cite{Wadati-1982})
		\begin{align}\label{WKI-transformation}
			\mathrm{d}X=-f\,\mathrm{d}x+\mathcal H\,\mathrm{d}t,\qquad T:=t,\qquad
			\hat q:=\mathrm{i}\frac{g}{f},\qquad \hat r:=\mathrm{i}\frac{e}{f}.
		\end{align}
		Equivalently, $X_x=-f$, $X_t=\mathcal H$, and therefore
		\begin{align}\label{WKI-derivatives}
			\partial_x=-f\partial_X,\qquad
			\partial_t=\partial_T+\mathcal H\partial_X.
		\end{align}
	Under the reciprocal transformation \eqref{WKI-transformation}, \eqref{HFM-1p1} becomes
	\begin{subequations}\label{WKI-system}
		\begin{align}
			\hat q_T=\pm\mathrm{i}\left(\frac{\hat q}{\sqrt{1-\hat q\hat r}}\right)_{XX}, \\
			\hat r_T=\mp\mathrm{i}\left(\frac{\hat r}{\sqrt{1-\hat q\hat r}}\right)_{XX}.
		\end{align}
	\end{subequations}
	The variables in \eqref{WKI-system} are expressed in terms of the ASDYM matrix $J$ as
	\begin{align}
		\hat q:=\frac{2\mathrm{i}J_{12}J_{22}}{J_{11}J_{22}+J_{12}J_{21}},\qquad \hat r:=-\frac{2\mathrm{i}J_{11}J_{21}}{J_{11}J_{22}+J_{12}J_{21}}.
	\end{align}
	Under the Hermitian reduction $S=S^\dagger$, one has $e=g^*$ and $\hat r=-\hat q^*$, and \eqref{WKI-system} reduces to the WKI equation \eqref{Shimizu-Wadati}:
	\begin{align}\label{WKI-final}
		\hat q_T=\pm\mathrm{i}\left(\frac{\hat q}{\sqrt{1+|\hat q|^2}}\right)_{XX}.
	\end{align}

Equation \eqref{WKI-system} can also be written directly in terms of the ASDYM matrices. Since $f^2+eg=1$, this implies
	\begin{align}\label{sqrt-identity}
		1-\hat q\hat r=1+\frac{eg}{f^2}=\frac{f^2+eg}{f^2}=\frac{1}{f^2},
		~~~~\text{i.e.}~~~~\sqrt{1-\hat q\hat r}=\frac{1}{f}.
	\end{align}
Changing this branch only reverses
the correlated signs in \eqref{WKI-system}. On a domain where $f\neq0$,
	\begin{align}
		\frac{\hat q}{\sqrt{1-\hat q\hat r}}=f\hat q=\mathrm ig,~~~~
		\frac{\hat r}{\sqrt{1-\hat q\hat r}}=f\hat r=\mathrm ie .
	\end{align}
Consequently, the WKI system \eqref{WKI-system} reduces to
	\begin{align}\label{WKI-system-S}
		\hat q_T=\mp\,\partial_X^2g,~~~~\hat r_T=\pm\,\partial_X^2e,
	\end{align}
Thus the evolution is determined by the off-diagonal entries of $S$. All quantities in the
transformation are explicit functions of the ASDYM matrix $J$:
	\begin{align}\label{WKI-in-J}
		&f=\frac{J_{11}J_{22}+J_{12}J_{21}}{\mathrm{det}(J)},~~~~
		g=\frac{2J_{12}J_{22}}{\mathrm{det}(J)},~~~~
		e=-\frac{2J_{11}J_{21}}{\mathrm{det}(J)},\\
		&\partial_X=-\frac{1}{f}\partial_x,~~~~
		\mathrm dX=-f\,\mathrm dx+\mathcal H\,\mathrm dt .
	\end{align}
Under the Hermitian condition $S=S^\dagger$ one has $e=g^*$ and $f\in\mathbb R$, so that
\eqref{WKI-system-S} reduces to the single equation $\hat q_T=\mp\partial_X^2g$ for
$\hat q=\mathrm ig/f$, which is \eqref{WKI-final}.

\subsection{Lax pairs of the reduced systems}\label{sec-3-3}

The derivation above is formulated entirely in terms of the ASDYM matrices and does not require Lax pairs.
The Lax pairs of the reduced systems can, however, be obtained from the ASDYM linear system under
the same constraint \eqref{dimensional-reduction}.

Recall that the ASD condition \eqref{ASD-condition-expand} is the compatibility condition of
\begin{align}\label{ASDYM-Lax}
	\mathcal L\Psi:=(\nabla_z-\zeta\nabla_{\tilde w})\Psi=0,\qquad
	\mathcal M\Psi:=(\nabla_w-\zeta\nabla_{\tilde z})\Psi=0,
\end{align}
where $\zeta$ is a spectral parameter, since
$[\mathcal L,\mathcal M]=F_{zw}-\zeta(F_{z\tilde z}-F_{w\tilde w})-\zeta^2F_{\tilde z\tilde w}$
vanishes for all $\zeta$ if and only if \eqref{ASD-condition-expand} holds.
In accordance with \eqref{dimensional-reduction-integrated}, the symmetry generated by
$\partial_{\tilde w}$ is lifted to the wave function by
\begin{align}\label{Psi-reduction}
	\Psi(z,\tilde z,w,\tilde w)=e^{-\tilde wC}\widehat\Psi(z,\tilde z,w),
	\qquad \partial_{\tilde w}\Psi=-C\Psi,
\end{align}
which is consistent because $C$ commutes with its exponential. The variable $\tilde w$ then
drops out of \eqref{ASDYM-Lax} identically.

\medskip
\noindent\textbf{(i) The $J$-gauge: the $(2+1)$-dimensional NLS system.}
In the $J$-gauge \eqref{gauge-potential-J} we have $A_{\tilde z}=A_{\tilde w}=0$ and,
by \eqref{Miura-transformation-J-K}, $A_z=\partial_{\tilde w}K=[K,C]$ and
$A_w=\partial_{\tilde z}K$. Hence \eqref{ASDYM-Lax} reduces to
\begin{subequations}\label{NLS-Lax}
	\begin{align}
		&\partial_z\widehat\Psi=-(\zeta C+\mathcal Q)\widehat\Psi,\qquad
		\mathcal Q:=[K,C]=\begin{pmatrix} 0 & q \\ r & 0\end{pmatrix},
		\label{NLS-Lax-a}\\
		&\partial_w\widehat\Psi=\zeta\,\partial_{\tilde z}\widehat\Psi
		-(\partial_{\tilde z}K)\widehat\Psi,\qquad
		\partial_{\tilde z}K=\begin{pmatrix}
			-\partial_z^{-1}(qr)_{\tilde z} & -q_{\tilde z}\\
			r_{\tilde z} & \partial_z^{-1}(qr)_{\tilde z}\end{pmatrix}.
		\label{NLS-Lax-b}
	\end{align}
\end{subequations}
Equation \eqref{NLS-Lax-a} is the Zakharov--Shabat (AKNS) spectral problem,
while \eqref{NLS-Lax-b} contains $\partial_{\tilde z}$ and has the characteristic form of a $(2+1)$-dimensional spectral problem. The zero-curvature condition of \eqref{NLS-Lax} reads
\begin{align}\label{zero-curvature}
	\partial_z(\partial_{\tilde z}K)-\partial_w\mathcal Q+\zeta\,\partial_{\tilde z}\mathcal Q
	+[\zeta C+\mathcal Q,\,\partial_{\tilde z}K]=0.
\end{align}
Its $O(\zeta)$ part, $\partial_{\tilde z}\mathcal Q+[C,\partial_{\tilde z}K]=0$, holds identically,
while its $O(\zeta^0)$ part gives the $(2+1)$-dimensional NLS system \eqref{NLS-system}.

\medskip
\noindent\textbf{(ii) The DNLS family.}
Put $\zeta=\lambda^2$ and set $\chi=G^{-1}\widehat\Psi$ with
\begin{align}\label{DNLS-gauge}
	G:=T\,\Delta\,D_\gamma,\qquad
	T=\begin{pmatrix}1&0\\ p&1\end{pmatrix},\qquad
	\Delta=\diag(\lambda^{-1/2},\lambda^{1/2}),\qquad
	D_\gamma=\diag(s^{\gamma/2},s^{-\gamma/2}),
\end{align}
where $p$ is defined by \eqref{p-def} and $s=J_{11}$ by \eqref{gauge-trans-1}.
Using the Miura relation \eqref{Miura-NLS-GI} in the form $r=-p_z+p^2q$ and
$\partial_z\ln s=-pq$, a direct computation turns \eqref{NLS-Lax-a} into
\begin{align}\label{DNLS-Lax}
	\begin{aligned}
		&\partial_z\chi=-\Big[\lambda^2C+\big(1-\tfrac{\gamma}{2}\big)uv\,\sigma_3
		+\lambda\big(vN^{+}-uN^{-}\big)\Big]\chi,\\[2mm]
		&N^{+}=\begin{pmatrix}0&1\\0&0\end{pmatrix},\qquad
		N^{-}=\begin{pmatrix}0&0\\1&0\end{pmatrix},
	\end{aligned}
\end{align}
with $(u,v)$ as in \eqref{gauge-trans-2}. This is the spectral problem of the
DNLS hierarchies: $\gamma=0$ gives the GI spectral problem,
$\gamma=1$ the CLL one, and $\gamma=2$ the KN one. The second operator follows from
\eqref{NLS-Lax-b} by the same conjugation,
\begin{align}
	\partial_w\chi=\lambda^2\partial_{\tilde z}\chi
	-G^{-1}\Big[\partial_wG-\lambda^2\partial_{\tilde z}G
	+(\partial_{\tilde z}K)G\Big]\chi.
\end{align}
In particular, the $(2+1)$-dimensional GDNLS system \eqref{GDNLS-system}, and hence the $(2+1)$-dimensional
GDNLS equation \eqref{GDNLS-2p1}, possesses a zero-curvature representation with a
spectral parameter, which establishes its integrability in the usual Lax-pair sense.

\medskip
\noindent\textbf{(iii) The $P$-gauge: the $(2+1)$-dimensional HF system.}
The gauge transformation relating the $J$- and $P$-gauges is $\Phi=J^{-1}\Psi$,
and since $J$ obeys \eqref{dimensional-reduction-integrated}, the lift \eqref{Psi-reduction}
is preserved. In the $P$-gauge \eqref{gauge-potential-C} we have $C_z=C_w=0$,
$C_{\tilde w}=\partial_zQ=C-\frac{1}{2}S$ by \eqref{Q-S}, and $C_{\tilde z}=\partial_wQ$
by \eqref{Q-w}, so that \eqref{NLS-Lax} becomes
\begin{subequations}\label{HF-Lax}
	\begin{align}
		&\partial_z\widehat\Phi=-\frac{\zeta}{2}\,S\,\widehat\Phi,
		\label{HF-Lax-a}\\
		&\partial_w\widehat\Phi=\zeta\Big[\partial_{\tilde z}\widehat\Phi
		+\Big(\tfrac{1}{4}[S,\partial_{\tilde z}S]+\kappa S\Big)\widehat\Phi\Big].
		\label{HF-Lax-b}
	\end{align}
\end{subequations}
This is a Landau--Lifshitz-type spectral problem coupled to a $(2+1)$-dimensional evolution equation.
Its zero-curvature condition reproduces \eqref{S-zt-evo} at order $\zeta^2$ and \eqref{3.28}
at order $\zeta$. These are the two relations from which Theorem~\ref{thm-1} was derived.

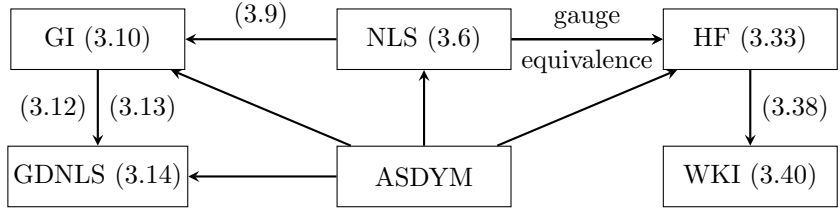
\begin{figure}[H]
		\centering
		\begin{tikzpicture}[
			box/.style={draw, rectangle, minimum width=2.3cm, minimum height=0.8cm, inner sep=2pt, font=\small},
			arrow/.style={->, >=stealth, thick},
			node distance=1cm and 1cm
			]

			\node[box] (NLS) at (0,0) {NLS \eqref{NLS-system}};
			\node[box, below=1cm of NLS] (ASDYM) {ASDYM};
			\node[box, left=2cm of NLS] (GI) {GI \eqref{GI-system}};
			\node[box, below=1cm of GI] (GDNLS) {GDNLS \eqref{GDNLS-system}};
			\node[box, right=2cm of NLS] (HF) {HF \eqref{Heisenberg-model}};
			\node[box, below=1cm of HF] (WKI) {WKI \eqref{WKI-system}};

			\draw[arrow] (NLS) --node[midway, above, sloped, font=\small] {gauge} (HF);
			\draw[arrow] (NLS) --node[midway, below, sloped, font=\small] {equivalence} (HF);
			\draw[arrow] (HF) --node[midway, right, font=\small] {\eqref{WKI-transformation}} (WKI);
			\draw[arrow] (NLS) --node[midway, above, font=\small] {\eqref{Miura-NLS-GI}} (GI);
			\draw[arrow] (GI) --node[midway, left, font=\small] {\eqref{gauge-trans-1}} node[midway, right, font=\small] {\eqref{gauge-trans-2}} (GDNLS);
			\draw[arrow] (ASDYM) -- (NLS);
			\draw[arrow] (ASDYM) -- (GI);
			\draw[arrow] (ASDYM) -- (GDNLS);
			\draw[arrow] (ASDYM) -- (HF);
		\end{tikzpicture}
		\caption{Integrable systems derived within the ASDYM reduction framework and their interrelations.
The NLS--HF link is the gauge equivalence described in \cite{Myrzakulov-1987,Zakharov-1979}, and the NLS--GI link is the Miura transformation \eqref{Miura-NLS-GI}. The GI--GDNLS and HF--WKI links are, respectively, the gauge and reciprocal transformations indicated in the diagram; the arrows from ASDYM denote reductions under \eqref{dimensional-reduction}.}
		\label{fig-1}
	\end{figure}

\medskip
\noindent\textbf{(iv) The reciprocal transformation: the WKI equation.}
For the $(1+1)$-dimensional reduction of Section~\ref{sec-3-2}, set $\partial_{\tilde z}=\partial_z$ (equivalently, $\partial_x=\partial_y$). After applying the time-dependent spatial translation described in Remark~\ref{rem-spatial-translation}, take $\kappa=0$ (equivalently, $u=0$). Eliminating $\partial_{\tilde z}\widehat\Phi$ from \eqref{HF-Lax-b} by means of \eqref{HF-Lax-a} then yields the classical Lax pair of the $(1+1)$-dimensional HF model
\eqref{HFM-1p1},
\begin{align}\label{HF-Lax-1p1}
	\partial_z\widehat\Phi=-\frac{\zeta}{2}S\widehat\Phi,\qquad
	\partial_w\widehat\Phi=\Big[-\frac{\zeta^2}{2}S
	+\frac{\zeta}{4}[S,\partial_zS]\Big]\widehat\Phi,
\end{align}
whose compatibility gives $\partial_wS=-\frac{1}{2}[S,\partial_z^2S]$, that is,
\eqref{HFM-1p1}. Passing to the real coordinates, where $\partial_z=\mathrm{i}\partial_x$, and
applying the reciprocal transformation \eqref{WKI-transformation}, for which
$\partial_x=-f\partial_X$ and
	$\partial_t=\partial_T+\mathcal H\partial_X$, the first equation of \eqref{HF-Lax-1p1} becomes
\begin{align}\label{WKI-Lax}
	\partial_X\widehat\Phi=-\frac{\mathrm{i}\zeta}{2}\,\frac{S}{f}\,\widehat\Phi
	=\begin{pmatrix}
		-\mathrm{i}\lambda & -\lambda\hat q \\
		-\lambda\hat r & \mathrm{i}\lambda
	\end{pmatrix}\widehat\Phi,
	\qquad \lambda:=\frac{\zeta}{2},
\end{align}
since $S/f=\left(\begin{smallmatrix}1&-\mathrm{i}\hat q\\ -\mathrm{i}\hat r&-1\end{smallmatrix}\right)$
by \eqref{WKI-transformation}. The second equation simultaneously becomes
\begin{align}\label{WKI-Lax-time}
	\partial_T\widehat\Phi=
		\left[\frac{\mathrm{i}\zeta^2}{2}S-
		\frac{\zeta f}{4}[S,S_X]+
		\frac{\mathrm{i}\zeta}{2}\frac{\mathcal H}{f}S\right]\widehat\Phi.
\end{align}
Equation \eqref{WKI-Lax} is the
WKI scattering problem, in which the spectral parameter multiplies the
potentials. Together with \eqref{WKI-Lax-time}, its compatibility gives
	\eqref{WKI-system} (cf.~\cite{Wadati-1982,Wadati-1979}). Thus the WKI branch of
Figure~\ref{fig-1} also inherits a Lax pair from \eqref{ASDYM-Lax}, and
Remark~\ref{rem-WKI-realization} in Section~\ref{sec-4-2} shows that the underlying symmetry
$\partial_x=\partial_y$ can be realized inside the Cauchy matrix scheme, so that explicit
solutions of \eqref{WKI-system} are available as well.

\medskip
Thus the classical gauge equivalence between the $(2+1)$-dimensional NLS and HF equations is realized by the ASDYM gauge transformation $g=J^{-1}$ connecting the $J$- and $P$-gauges. All four Lax
pairs above descend from the single linear system \eqref{ASDYM-Lax}.

The relationships among the reduced integrable systems are summarized in Figure~\ref{fig-1}.

Although the above figure describes the ASDYM reduction framework in $\mathrm{GL}(2,\mathbb C)$,
there is no essential change to extend the research to
$\mathrm{GL}(M,\mathbb C)$ at the matrix level.
Such extensions will be sketched in Appendix \ref{sec-3-4}.

\section{Special solutions}\label{sec-4}

We use the Cauchy matrix approach to realize the constraint \eqref{dimensional-reduction} and to
construct special solutions of the ASDYM equations and their reductions.

	\subsection{Cauchy matrix structure}\label{sec-4-1}
	The Cauchy matrix approach to the ASDYM equations was developed in \cite{LSS-2022,LSS-2023} as a discrete-parameter analogue of the direct linearization framework \cite{LSS-DL-2025}. With a nontrivial seed, it is also closely related to a vectorial binary Darboux transformation \cite{Dimakis-2013,LSS-PRD-2025}.

	For $G=\mathrm{GL}(2,\mathbb C)$, the Cauchy matrix structure used below is summarized by the following theorem (cf.~Theorem~3.3 of Ref.~\cite{LSS-PRD-2025}).
	\begin{theorem}\label{thm-CMA}
		Assume that $\Xi,\Lambda,\Omega,\eta$, and $\theta$ satisfy the following Cauchy matrix structure:
		\begin{subequations}\label{Cauchy-matrix-structure}
			\begin{align}
				&\text{the Sylvester equation:} \qquad \Xi\Omega-\Omega\Lambda=\eta\theta,
\label{Cauchy-matrix-structure-SE}\\
				&\text{dispersion relations:} \qquad
				\begin{cases}
					&\partial_z\eta=\Xi(\partial_{\tilde w}\eta),\qquad \partial_z\theta=(\partial_{\tilde w}\theta)\Lambda, \\
					&\partial_w\eta=\Xi(\partial_{\tilde z}\eta),\qquad \partial_w\theta=(\partial_{\tilde z}\theta)\Lambda,
				\end{cases} \label{4.1b}\\
				&\text{master function:} \qquad S^{(i,j)}:=\theta\Lambda^j\Omega^{-1}\Xi^i\eta. \label{Sij}
			\end{align}
		\end{subequations}
		Here $\Xi,\Lambda\in\mathbb C^{N\times N}$ are constant spectral-parameter matrices, independent of $(z,\tilde z,w,\tilde w)$. On a domain $U\subset\mathbb C^4$, the plane-wave matrices satisfy $\eta\in\operatorname{Mat}_{N\times2}(\mathcal O(U))$ and $\theta\in\operatorname{Mat}_{2\times N}(\mathcal O(U))$, while $\Omega\in\operatorname{Mat}_{N\times N}(\mathcal O(U))$ is the Cauchy-type matrix determined by the Sylvester equation \eqref{Cauchy-matrix-structure-SE}. Consequently, $S^{(i,j)}\in\operatorname{Mat}_{2\times2}(\mathcal O(U))$.

	We assume that $\Xi$ and $\Lambda$ are invertible, that they do not share any eigenvalues,
	and that $\Omega$ is invertible on the domain under consideration. The invertibility assumptions are needed to define the negative powers and $\Omega^{-1}$ in
	\eqref{Sij}, while the disjoint-spectrum assumption ensures that $\Omega$ is uniquely determined by
	\eqref{Cauchy-matrix-structure-SE} for given $\Xi$, $\Lambda$, $\eta$, and $\theta$ \cite{Sylvester-1884}.

		Then $S^{(i,j)}$ satisfies the difference-to-product relation
		\begin{align}\label{diff-to-prod}
			S^{(i+1,j)}-S^{(i,j+1)}=S^{(0,j)}S^{(i,0)},
		\end{align}
		and the following evolution:
		\begin{subequations}\label{Sij-evo}
			\begin{align}
				\label{Sij-evo-1}&\partial_wS^{(i-1,j)}-\partial_{\tilde z}S^{(i,j)}+\partial_{\tilde z}S^{(0,j)}S^{(i-1,0)}=0, \\
				\label{Sij-evo-2}&\partial_zS^{(i-1,j)}-\partial_{\tilde w}S^{(i,j)}+\partial_{\tilde w}S^{(0,j)}S^{(i-1,0)}=0.
			\end{align}
		\end{subequations}
		The corresponding Cauchy-matrix solutions are given by
		\begin{subequations}\label{CMA-solution}
			\begin{align}
				&J=I-S^{(-1,0)}=I-\theta\Omega^{-1}\Xi^{-1}\eta,\qquad
K=S^{(0,0)}=\theta\Omega^{-1}\eta, \label{CMA-solution-a} \\
				&P=I+S^{(0,-1)}=I+\theta\Lambda^{-1}\Omega^{-1}\eta,\qquad
Q=-S^{(-1,-1)}=-\theta\Lambda^{-1}\Omega^{-1}\Xi^{-1}\eta.
			\end{align}
		\end{subequations}
		They also satisfy
		\begin{subequations}
			\begin{align}
				&\det(J)=\frac{\det(\Lambda)}{\det(\Xi)},\qquad
\tr(K)=\tr(\Xi)-\tr(\Lambda), \label{Det-J-Tr-K}\\
				&\det(P)=\frac{\det(\Xi)}{\det(\Lambda)},\qquad
\tr(Q)=\tr(\Lambda^{-1})-\tr(\Xi^{-1}).
			\end{align}
		\end{subequations}
	\end{theorem}
	\begin{proof}
		The difference-to-product formula \eqref{diff-to-prod} was proved in Eq.~(22) of Ref.~\cite{LSS-2022} and Eq.~(2.9) of Ref.~\cite{LSS-2023}. The evolution relations for $S^{(i,j)}$ were proved in Theorem~2 of Ref.~\cite{LSS-DL-2025}. The gauge-potential relations \eqref{Miura-transformation-J-K} and \eqref{Miura-transformation-P-Q} then follow directly from \eqref{Sij-evo} and \eqref{CMA-solution}.
		For completeness, we give a detailed proof in Appendix~\ref{app-1}.
	\end{proof}

	If $\Xi$ and $\Lambda$ are diagonal, the Cauchy matrix structure \eqref{Cauchy-matrix-structure} is realized by (cf.~Eq.~(2.6) of Ref.~\cite{Ohta-2024} or Eq.~(5.9) of Ref.~\cite{LMZ-2026}):
	\begin{subequations}
		\begin{align}
			&\Xi=\diag(\xi_1,\ldots,\xi_N),\qquad \eta=(\phi_{is}(\xi_iz+\tilde w,\xi_iw+\tilde z))_{N\times 2}, \\
			&\Lambda=\diag(\lambda_1,\ldots,\lambda_N),\qquad \theta=(\psi_{sj}(\lambda_jz+\tilde w,\lambda_jw+\tilde z))_{2\times N}, \\
			&\Omega=(\Omega_{ij})_{N\times N},\qquad \Omega_{ij}=\frac{\sum_{s=1}^2\phi_{is}\psi_{sj}}{\xi_i-\lambda_j},
		\end{align}
	\end{subequations}
	where $\xi_i, \lambda_i \in \mathbb C$, $\phi_{is}$ and $\psi_{sj}$ are arbitrary holomorphic functions.
	Write $\eta$ and $\theta$ in block form as
	\begin{align}
		\eta=
		\begin{pmatrix}
			\eta_1 & \eta_2
		\end{pmatrix},\qquad
		\theta=
		\begin{pmatrix}
			\theta_1 \\
			\theta_2
		\end{pmatrix}.
	\end{align}
	The entries of $J$ are
	\begin{subequations}\label{J-expand}
		\begin{align}
			&J_{11}=1-\theta_1\Omega^{-1}\Xi^{-1}\eta_1,\qquad
			J_{12}=-\theta_1\Omega^{-1}\Xi^{-1}\eta_2, \\
			&J_{21}=-\theta_2\Omega^{-1}\Xi^{-1}\eta_1,\qquad
			J_{22}=1- \theta_2\Omega^{-1}\Xi^{-1}\eta_2,
		\end{align}
	\end{subequations}
	while those of $K$ are
	\begin{subequations}\label{K-expand}
		\begin{align}
			&K_{11}=\theta_1\Omega^{-1}\eta_1,\qquad K_{12}=\theta_1\Omega^{-1}\eta_2, \\
			&K_{21}=\theta_2\Omega^{-1}\eta_1,\qquad K_{22}=\theta_2\Omega^{-1}\eta_2.
		\end{align}
	\end{subequations}
	This parameterization by the holomorphic functions $\phi_{is}$ and $\psi_{sj}$ gives ASDYM solutions with
$\mathfrak{sl}(2,\mathbb C)$-valued gauge potentials.\footnote{Indeed, the Cauchy matrix
structure forces $\det(J)$ and $\tr(K)$ to be constants, see \eqref{Det-J-Tr-K},
which, by Remark~\ref{rem-SL2-SU2}(i), makes all $A_\alpha$ traceless. By Remark~\ref{rem-SL2-SU2}(ii), the normalization \eqref{normalization-SU2} then puts the normalized $J$ in
$\mathrm{SL}(2,\mathbb C)$. Note that $\det(J)=\det(\Lambda)/\det(\Xi)$
and $\tr(K)=\tr(\Xi)-\tr(\Lambda)$ are not equal to $1$ and $0$
in general.} The corresponding reduced fields are then obtained from the entries of $J$ and $K$ after imposing the dimensional reduction condition.

\subsection{Realization of the constraint and reduced Lagrangian density}\label{sec-4-2}

In this subsection, we work at the $\mathrm{SL}(2,\mathbb C)$ group level in the sense of
Remark~\ref{rem-SL2-SU2}(ii), with $(z,\tilde z,w,\tilde w)$ regarded as complex coordinates.

To impose the dimensional reduction condition \eqref{dimensional-reduction}, require the following $\tilde w$-dependence of $\eta$ and $\theta$ (cf.~Eq.~(5.12) of Ref.~\cite{LMZ-2026}):
	\begin{align}\label{4.10}
		\partial_{\tilde w}\eta=\eta C,\qquad \partial_{\tilde w}\theta=-C\theta.
	\end{align}
Differentiating the Sylvester equation \eqref{Cauchy-matrix-structure-SE} with respect to $\tilde w$ and using \eqref{4.10} gives
\[\Xi (\partial_{\tilde w}\Omega)-(\partial_{\tilde w}\Omega)\Lambda
=(\eta_{\tilde w})\theta + \eta (\theta_{\tilde w}).\]
Thus $\Xi(\partial_{\tilde w}\Omega)-(\partial_{\tilde w}\Omega)\Lambda=0$, and uniqueness of the Sylvester equation gives
\begin{equation}\label{4.11}
\partial_{\tilde w}\Omega=0.
\end{equation}
Differentiating $S^{(i,j)}$ in \eqref{Sij} with respect to $\tilde w$ and using \eqref{4.10} and \eqref{4.11} gives
\begin{align}\label{4.12}
		\partial_{\tilde w}S^{(i,j)}=-CS^{(i,j)}+S^{(i,j)}C=[S^{(i,j)},C].
	\end{align}
Thus $S^{(i,j)}$ satisfies the same reduction constraint. Since $J,K,P$, and $Q$ are expressed in terms of $S^{(i,j)}$ through \eqref{CMA-solution}, we obtain the following proposition.

\begin{proposition}\label{prop-1}
If \eqref{4.10} holds in the Cauchy matrix framework of Theorem~\ref{thm-CMA}, then the matrices $J,K,P,Q$ satisfy \eqref{dimensional-reduction}.
\end{proposition}

Under \eqref{4.10}, the plane-wave factors reduce to
	\begin{subequations}\label{phi-psi}
		\begin{align}
			&\phi_{i1}=\exp\left(\frac{1}{2}(\xi_iz+\tilde w)\right)\varrho_{i1}\left(\xi_iw+\tilde z\right), \\
			&\phi_{i2}=\exp\left(-\frac{1}{2}(\xi_iz+\tilde w)\right)\varrho_{i2}\left(\xi_iw+\tilde z\right), \\
			&\psi_{1j}=\exp\left(-\frac{1}{2}(\lambda_jz+\tilde w)\right)\varsigma_{1j}\left(\lambda_jw+\tilde z\right), \\
			&\psi_{2j}=\exp\left(\frac{1}{2}(\lambda_jz+\tilde w)\right)\varsigma_{2j}\left(\lambda_jw+\tilde z\right),
		\end{align}
	\end{subequations}
	where $\varrho_{is}$ and $\varsigma_{sj}$ ($i,j\in\{1,\ldots,N\}$ and $s\in\{1,2\}$) are arbitrary holomorphic functions.

The same reduction also gives a direct $\tau$-function expression for the ASDYM Lagrangian density.

\begin{theorem}\label{thm-tau-Lagrangian}
Under the assumptions of Theorem~\ref{thm-CMA} and the additional constraint \eqref{4.10}, the
ASD Yang--Mills Lagrangian density can be written as
    \begin{align}\label{L-tau-theorem}
        \mathcal L_{[\mathrm{ASD}]}=4\partial^2_{\tilde z}\partial^2_z(\ln\tau),\qquad \tau:=\det(\Omega).
    \end{align}
\end{theorem}

\begin{proof}
	Under the constraint \eqref{4.10}, the first line of \eqref{4.1b} reduces to
	\begin{align*}
		\partial_z\eta=\Xi\eta C,\qquad \partial_z\theta=-C\theta\Lambda.
	\end{align*}
	Therefore,
	\begin{align*}
		\Xi(\partial_z\Omega)-(\partial_z\Omega)\Lambda
		=(\partial_z\eta)\theta+\eta(\partial_z\theta)=\Xi(\eta C\theta)-(\eta C\theta)\Lambda,
	\end{align*}
	that is,
	\begin{align*}
		\Xi(\partial_z\Omega-\eta C\theta)-(\partial_z\Omega-\eta C\theta)\Lambda=0.
	\end{align*}
	By uniqueness of the Sylvester equation, one has
	\begin{equation}\label{4.15}
		\partial_z\Omega=\eta C\theta.
	\end{equation}
	Using \eqref{3.3a},
	\begin{align*}
		\partial_z(K_{11})&=\frac{1}{2}\partial_z(K_{11}-K_{22})
		=\partial_z\tr(\theta\Omega^{-1}\eta C)
		=\partial_z\tr(\Omega^{-1}\eta C\theta),
	\end{align*}
	where $C=\frac{1}{2}\sigma_3$ and we used the cyclicity of the trace, $\tr(AB)=\tr(BA)$.
	Using \eqref{4.15} gives
	\begin{align}
		\partial_z(K_{11})=\partial_z\tr(\Omega^{-1}\Omega_z)
		=\partial_z^2(\ln\det(\Omega))=\partial_z^2(\ln\tau). \label{bilinear-rep}
	\end{align}
	Next, recall the Lagrangian density formula \eqref{L-K}, namely
	\[
	\mathcal L_{[\mathrm{ASD}]}
	=4\tr\left((\partial_{\tilde z}^2K)(\partial_{\tilde w}^2K)
	-(\partial_{\tilde z\tilde w}K)(\partial_{\tilde z\tilde w}K)\right)=:\mathcal L_{[K]},
	\]
	and the constraint  $\partial_{\tilde w}K=[K,C]$ in \eqref{dimensional-reduction}.
	Substitution of $\partial_{\tilde w}K=[K,C]$ into \eqref{L-K} gives
	\begin{align*}
		\mathcal L_{[K]}=4\tr\left((\partial_{\tilde z}^2K)[[K,C],C]-([\partial_{\tilde z}K,C])^2\right)
		=4\partial^2_{\tilde z}(K_{12}K_{21})=4\partial_{\tilde z}^2\partial_z(K_{11}).
	\end{align*}
	Combining this expression with \eqref{bilinear-rep} yields
	\begin{align}
		\mathcal L_{[K]}=4\partial^2_{\tilde z}\partial^2_z(\ln\tau),
	\end{align}
	which proves the theorem.

\end{proof}

The $\tau$-function in Theorem~\ref{thm-tau-Lagrangian} reduces to a standard Grammian
$\tau$-function of the KP equation when $\varrho_{is}$ and $\varsigma_{sj}$ are specialized to exponential functions:
	\begin{subequations}
		\begin{align}
			&\varrho_{i1}=A_{i1}\exp\left(\frac{1}{2}(\xi_i^2\tilde z+\xi_i^3 w)\right),\qquad \varrho_{i2}=A_{i2}\exp\left(-\frac{1}{2}(\xi_i^2\tilde z+\xi_i^3 w)\right), \\
			&\varsigma_{1j}=B_{1j}\exp\left(-\frac{1}{2}(\lambda_j^2\tilde z+\lambda_j^3 w)\right),\qquad \varsigma_{2j}=B_{2j}\exp\left(\frac{1}{2}(\lambda_j^2\tilde z+\lambda_j^3 w)\right).
		\end{align}
	\end{subequations}
These yield
	\begin{subequations}
		\begin{align}
			&\Omega_{ij}=\frac{C_{ij}\exp(H_i-\Theta_j)+D_{ij}\exp(-H_i+\Theta_j)}{\xi_i-\lambda_j},
			\label{Omega-Cauchy-exp}\\
			&H_i=\frac{1}{2}\left(\xi_i z+\xi_i^2\tilde z+\xi_i^3w\right), \qquad C_{ij}=A_{i1}B_{1j}, \\
			&\Theta_j=\frac{1}{2}\left(\lambda_jz+\lambda_j^2\tilde z+\lambda_j^3w\right),\qquad D_{ij}=A_{i2}B_{2j}.
		\end{align}
	\end{subequations}
With this specialization, $\tau=\det(\Omega)$ is a Grammian $\tau$-function for the KP equation (e.g., \cite{Matveev-book}):
	\begin{align}\label{KP-eq}
  	  (-4u_w+u_{zzz}+6uu_z)_z+3u_{\tilde z\tilde z}=0,\qquad u=2(\ln\tau)_{zz}.
	\end{align}
\begin{remark}
	Here \eqref{KP-eq} is understood in the complex variables $(z,\tilde z,w)$, and
	$u=2\big(\ln\det(\Omega)\big)_{zz}$ solves the KP equation \eqref{KP-eq}.
	If $(z,\tilde z,w)\in\mathbb R^3$ and $\xi_i,\lambda_j\in\mathbb R$, the same formula gives Grammian-type line-soliton solutions of the KP-II equation
	\cite{Chakravarty-2009,Biondini-2006,Chakravarty-2008b}.
	If instead $\tilde z\in\mathrm{i}\mathbb R$ while $z,w\in\mathbb R$, the sign of the term
	$3u_{\tilde z\tilde z}$ is reversed and \eqref{KP-eq} takes the form of the KP-I equation.
	The last statement concerns the form of the equation only, we do not claim that the
	corresponding $\tau=\det(\Omega)$ then yields nonsingular solutions.
\end{remark}

\begin{remark}\label{rem-WKI-realization}
The $(1+1)$-dimensional symmetry $\partial_x=\partial_y$, which was imposed in Section~\ref{sec-3-2}
in order to reach the WKI equation, can also be realized inside the present Cauchy matrix scheme.
Since $\partial_x=\partial_y$ is equivalent to $\partial_z=\partial_{\tilde z}$, it suffices to
specialize the arbitrary holomorphic functions in \eqref{phi-psi} to
\begin{align*}
&\varrho_{i1}=A_{i1}\exp\Big(\tfrac{1}{2}\xi_i(\xi_iw+\tilde z)\Big),\qquad
\varrho_{i2}=A_{i2}\exp\Big(-\tfrac{1}{2}\xi_i(\xi_iw+\tilde z)\Big), \\
&\varsigma_{1j}=B_{1j}\exp\Big(-\tfrac{1}{2}\lambda_j(\lambda_jw+\tilde z)\Big),\qquad
\varsigma_{2j}=B_{2j}\exp\Big(\tfrac{1}{2}\lambda_j(\lambda_jw+\tilde z)\Big),
\end{align*}
for which \eqref{phi-psi} becomes
\[
\phi_{i1}=A_{i1}e^{\frac{1}{2}(\xi_i(z+\tilde z)+\xi_i^2w+\tilde w)},\qquad
\phi_{i2}=A_{i2}e^{-\frac{1}{2}(\xi_i(z+\tilde z)+\xi_i^2w+\tilde w)},
\]
\[
\psi_{1j}=B_{1j}e^{-\frac{1}{2}(\lambda_j(z+\tilde z)+\lambda_j^2w+\tilde w)},\qquad
\psi_{2j}=B_{2j}e^{\frac{1}{2}(\lambda_j(z+\tilde z)+\lambda_j^2w+\tilde w)}.
\]
Hence $\Omega,\eta,\theta$, and therefore $J$ and $K$, depend on $(z,\tilde z)$ only through
the combination $z+\tilde z$, so that $\partial_z=\partial_{\tilde z}$ holds identically.
The matrix $S=J^{-1}\sigma_3J$ built from such a $J$ then solves the $(1+1)$-dimensional HF model \eqref{HFM-1p1} (up to the time-dependent spatial translation described in Remark~\ref{rem-spatial-translation}); through the reciprocal transformation \eqref{WKI-transformation}, it therefore provides
solutions of the WKI equation \eqref{Shimizu-Wadati}.
This specialization is compatible with the unitary constraints \eqref{conj-constraints}, since
$\theta=\eta^\dagger$ amounts to $\lambda_j=\xi_j^*$ and $B_{sj}=A_{js}^*$.
\end{remark}

\subsection{Realization of the gauge conditions}\label{sec-4-3}

	The $J$- and $K$-matrices obtained from the Cauchy matrix approach have constant
$\det(J)$ and $\tr(K)$, so by Remark~\ref{rem-SL2-SU2}(i) the associated gauge
potentials are $\mathfrak{sl}(2,\mathbb C)$-valued. The matrix $J$ has not yet been normalized to
$\mathrm{SL}(2,\mathbb C)$: by \eqref{Det-J-Tr-K} together with the constraints \eqref{conj-constraints} below,
$\det(J)=\det(\Xi^*)/\det(\Xi)$ and
$\tr(K)=2\mathrm{i}\,\operatorname{Im}\,\tr(\Xi)$, which in general differ from $1$ and
$0$ (already for $N=1$ with $\xi_1=a+\mathrm{i}b$ one gets $\det(J)=\xi_1^*/\xi_1$ and
$\tr(K)=2\mathrm{i}b$). We therefore perform the normalization
\eqref{normalization-SU2}, after which $\det(J)=1$ and $\tr(K)=0$, i.e.,
$J\in\mathrm{SL}(2,\mathbb C)$, in accordance with Remark~\ref{rem-SL2-SU2}(ii).
The unitary conditions below complete step (iii) and give $J\in\mathrm{SU}(2)$.
	To impose the unitary conditions $K=-K^\dagger$ and $J^\dagger J=I$, require
	\begin{align}\label{conj-constraints}
		\Lambda=\Xi^\dagger,\qquad \theta=\eta^\dagger.
	\end{align}
The relation $\Omega=-\Omega^\dagger$ then follows from uniqueness. Indeed, taking the Hermitian conjugate of the
Sylvester equation \eqref{Cauchy-matrix-structure-SE} shows that $-\Omega^\dagger$ satisfies the same
Sylvester equation as $\Omega$.
	Under these constraints, we have
	\begin{subequations}
		\begin{align}
			&K^\dagger=(\theta\Omega^{-1}\eta)^\dagger=-\theta\Omega^{-1}\eta=-K, \\
			 &J^\dagger=(I-\theta\Omega^{-1}\Xi^{-1}\eta)^\dagger=I+\theta\Lambda^{-1}\Omega^{-1}\eta=P=J^{-1}.
		\end{align}
	\end{subequations}
	The relation $K=-K^\dagger$ also implies $K_{21}=-K_{12}^*$, which yields the conjugate reduction of
the $(2+1)$-dimensional NLS equation. The condition $J^\dagger=J^{-1}$ gives the corresponding Hermitian reduction of the $(2+1)$-dimensional HF equation:
	\begin{align}
		S^\dagger=(J^{-1}\sigma_3 J)^\dagger=J^{-1}\sigma_3 J=S.
	\end{align}
	The $(2+1)$-dimensional NLS and HF equations are known to be gauge equivalent.
In our construction, both systems arise from the same $\mathrm{SU}(2)$ ASDYM solution (in the
sense of Remark~\ref{rem-SL2-SU2}(iii)) written in two different gauges. Since the Lagrangian density \eqref{Yang--Mills-Lagrangian} is gauge invariant by \eqref{L-gauge-equivalent}, the two systems have the same Lagrangian density.
After the coordinate transformation $(x,y,t)=(\mathrm{i}z,\mathrm{i}\tilde z,\mathrm{i}w)$, their common Lagrangian density is
	\begin{align}\label{L-tau}
		\mathcal L=4\partial_y^2\partial_x^2(\ln\tau),\qquad \tau=\det(\Omega),
	\end{align}
	where the entries in $\Omega$ can be rewritten as
	\begin{align}\label{Omega-real}
		 \Omega_{ij}=\frac{1}{\xi_i-\xi_j^*}\left\{\exp\left[-\frac{\mathrm{i}(\xi_i-\xi_j^*)}{2}x\right]\rho_{i1}\rho_{j1}^*+\exp\left[\frac{\mathrm{i}(\xi_i-\xi_j^*)}{2}x\right]\rho_{i2}\rho_{j2}^*\right\}.
	\end{align}
	The holomorphic factors $\rho_{i1}$ and $\rho_{i2}$ can be written as
	\begin{subequations}\label{rho-holomorphic}
		\begin{align}
			\rho_{i1}:=\varrho_{i1}\left(\xi_iw+\tilde z\right)=\varrho_{i1}\left(-\mathrm{i}(y+\xi_it)\right), \\
			\rho_{i2}:=\varrho_{i2}\left(\xi_iw+\tilde z\right)=\varrho_{i2}\left(-\mathrm{i}(y+\xi_it)\right).
		\end{align}
	\end{subequations}
The factor $-\mathrm{i}$ in the argument can be absorbed into the holomorphic functions. We therefore regard $\rho_{i1}$ and $\rho_{i2}$ below as holomorphic functions of $y+\xi_i t$.

	We illustrate the $N=1$ one-soliton sector with the Zakharov system \eqref{Zakharov-system}.

	Set $\xi_1=\xi=a+\mathrm{i}b$, where $a,b\in\mathbb R$ and $b\neq0$, and write $F:=\rho_{11}$ and $H:=\rho_{12}$. Thus $F$ and $H$ are scalar holomorphic functions of $y+\xi t$. For $N=1$, $\Omega$ is scalar:
	\begin{align}
		\Omega=\frac{1}{2\mathrm{i}b}\left[\exp(bx)|F|^2+\exp(-bx)|H|^2\right].
	\end{align}
	A direct calculation gives
	\begin{align}
		|q|^2&=qr=-K_{12}K_{21}=\partial_x^2(\ln\det(\Omega)) \notag  \\
		 &=b^2\left\{1-\left[\frac{\exp(bx)|\varphi|-\exp(-bx)|\varphi|^{-1}}{\exp(bx)|\varphi|+\exp(-bx)|\varphi|^{-1}}\right]^2\right\} \notag \\
		&=b^2\sech^2(bx+\ln|\varphi|),
	\label{one-soliton-q}\end{align}
	where, on a domain on which $F H\neq0$, $\varphi(y+\xi t)=F/H$ is a nonvanishing
	holomorphic function.

	Different choices of $\varphi$ give different soliton profiles.
	\begin{enumerate}
		\item \emph{Line soliton.} Choose
		\begin{align}
			\varphi=\exp(\nu (y+\xi t))=\exp(\nu y+\nu\xi t),
		\end{align}
		where $\nu=c+\mathrm{i}d$ with $c,d\in\mathbb R$. Then
		\begin{align}
			|q|^2=b^2\sech^2\left[bx+cy+(ac-bd)t\right].
		\end{align}
		The maximum is attained on the plane
		\begin{align}
			bx+cy+(ac-bd)t=0.
		\end{align}
		\item \emph{Curved soliton.} Following \cite{Strachan-1992-3}, choose
		\begin{align}
			\varphi=\exp(\mu (y+\xi t)),
		\end{align}
			where $\mu(\cdot)$ is an arbitrary holomorphic function of its argument, distinct from the
			constant $\nu$ in the previous item. Thus $\mu=\mu(y+\xi t)$ is holomorphic in $(y,t)$.
		Then $|q|^2$ takes the form
		\begin{align}\label{curved-formulation}
			|q|^2=b^2\sech^2\left(bx+\frac{\mu+\mu^*}{2}\right).
		\end{align}
		For example, if $\mu=(y+\xi t)^2$, the maximum lies on the quadratic curve
		\begin{align}
			bx+y^2+2ayt+(a^2-b^2)t^2=0.
		\end{align}
		\item \emph{V-shaped soliton.} For a V-shaped profile in \eqref{curved-formulation}, take
		\begin{align}
			\mu=\ln(\cosh(y+\xi t)),
		\end{align}
		on a simply connected domain avoiding the zeros of $\cosh(y+\xi t)$.
		The soliton maximum lies on the curve
		\begin{align}
			2bx+\ln(\cosh(2y+2at)+\cos(2bt))=\ln 2.
		\end{align}
		The V-shaped asymptotics follow from
		$\ln\cosh x=|x|-\ln2+O(e^{-2|x|})$ as $|x|\to\infty$, while $\cos(2bt)$ remains bounded.
	\end{enumerate}
	For the above one-soliton solutions, \eqref{L-tau} becomes
	\begin{subequations}
		\begin{align}
			&\mathcal L=4b^2\partial_y^2(\sech^2X),\qquad X(x,y,t)=bx+\frac{1}{2}\left(\mu(y,t)+\mu^*(y,t)\right), \\
			&\mathcal L=4b^2\left[(4\,\sech^2X-6\,\sech^4X)(X_y)^2-2\,\sech^2X\tanh X(X_{yy})\right].
		\end{align}
	\end{subequations}
	For the line soliton with $\mu(y,t)=\nu y+\nu\xi t$, where $\nu=c+\mathrm{i}d$,
	\begin{align}
		\label{L-line-soliton}\mathcal L=8b^2c^2(2\,\sech^2X-3\,\sech^4X),
	\end{align}
	which agrees with Eq.~(3.44) of Ref.~\cite{Hamanaka-2020}.

\begin{figure*}[t]
	\centering
	\includegraphics[width=0.95\textwidth]{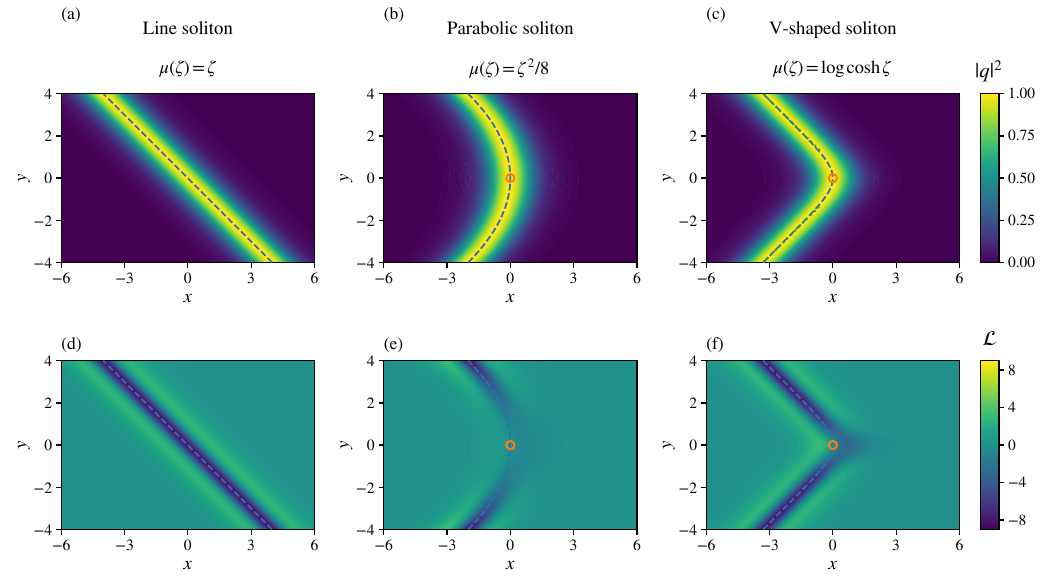}
	\caption{
		Intensity profiles $|q|^2$ (top row) and the corresponding common ASDYM
		Lagrangian densities $\mathcal L$ (bottom row) for the line, parabolic,
		and V-shaped one-soliton solutions.
		The parameters are $a=0$, $b=1$, $c=1$, $d=0$ and $t=0$, with
		$\mu(\zeta)=\zeta$, $\zeta^2/8$, and $\log\cosh\zeta$, respectively,
		where $\zeta=y+\mathrm{i}t$.
		Dashed curves indicate the intensity ridges $X=0$ in both rows.
	}
	\label{fig:one-soliton-profiles}
\end{figure*}

Having established the $\mathrm{SU}(2)$ reduction and the common Lagrangian density,
we now illustrate the one-soliton sector in the case $N=1$.
For the one-soliton solutions,
\[
|q|^2=b^2\sech^2 X, \qquad
\mathcal L=4b^2\partial_y^2(\sech^2 X) \qquad
X(x,y,t)=bx+\frac{1}{2}\bigl(\mu(y,t)+\mu^*(y,t)\bigr),
\]
so different choices of the holomorphic function $\mu$ determine different ridge
geometries.
Figure~\ref{fig:one-soliton-profiles} compares these quantities for the line,
parabolic, and V-shaped one-soliton solutions.

\section{Concluding remarks}\label{sec-5}

We have derived several classical $(2+1)$-dimensional integrable systems and their soliton solutions within a single $\mathrm{GL}(2,\mathbb C)$ ASDYM reduction framework. All reductions follow from imposing \eqref{dimensional-reduction} on the ASDYM matrices $J$, $K$, $P$, and $Q$, and the same constraint is realized in the Cauchy matrix formulation. The associated Lax pairs descend from a single ASDYM linear system. Under the unitary reduction, the $(2+1)$-dimensional NLS and HF systems arise from the same $\mathrm{SU}(2)$ ASDYM solution written in two different gauges. The reduced fields and gauge transformations are expressed explicitly in terms of the ASDYM matrices, and their relations are summarized in Figure~\ref{fig-1}.
Some results based on the $G=\mathrm{GL}(M,\mathbb C)$ ASDYM reduction framework
are presented in Appendix \ref{sec-3-4}.

The Cauchy matrix structure also gives the Lagrangian density directly in terms of the $\tau$-function
(see Theorem~\ref{thm-tau-Lagrangian}). More precisely,
\[
 \mathcal L_{[\mathrm{ASD}]}=4\partial_{\tilde z}^{2}\partial_z^{2}\ln\tau,
 \qquad \tau=\det(\Omega).
\]
Thus the same $\tau$-function determines the reduced fields, their Cauchy matrix solutions, and a
gauge-invariant density. For exponential plane wave factors, it becomes a Grammian-type KP $\tau$-function.
A detailed analysis of the anti-self-dual Yang--Mills action in this formulation is left for future work.

The Zakharov and Myrzakulov-I equations are gauge equivalent. Similarly, the DS system
	\begin{subequations}
		\begin{align}
			&\mathrm{i}q_t-q_{xx}+q_{yy}+2\phi q=0, \\
			&\phi_{xx}+\phi_{yy}+\delta(|q|^2)_{xx}-\delta(|q|^2)_{yy}=0,\qquad \delta=\pm1,
		\end{align}
	\end{subequations}
is gauge equivalent to the Ishimori equation \cite{Ishimori-1984,Leo-1992}:
	\begin{subequations}
		\begin{align}
			&\mathbf S_t+\mathbf S\times(\mathbf S_{xx}-\mathbf S_{yy})+\phi_x\mathbf S_y+\phi_y\mathbf S_x=0, \\
			&\phi_{xx}+\phi_{yy}-2\mathbf S\cdot(\mathbf S_x\times\mathbf S_y)=0.
		\end{align}
	\end{subequations}
	Whether the DS and Ishimori equations admit a parallel realization within the same ASDYM reduction framework remains open.
The GDNLS family is also related to the mixed NLS equation
	\begin{align}
		\mathrm{i}q_t+q_{xx}+\mathrm{i}\alpha(|q|^2q)_x+\beta|q|^2q=0,\qquad \alpha,\beta\in\mathbb R.
	\end{align}
This mixed NLS equation is gauge-related to \eqref{GDNLS} \cite{Kundu-1984,Kakei-1995}. Its direct realization in the ASDYM matrix formulation remains open.

\bigskip

\section*{Acknowledgments}
	The authors are grateful for valuable comments from Masashi Hamanaka, Yi C. Huang, and Kouichi Toda.
	The work of SSL is supported by the JSPS Grant (P25326) and the NSFC Grant (1261462).
	The work of KIM is supported by the JSPS Grant (JP22K03441, 23K22407) and the JSPS Bilateral Program (JPJSBP120247414).
	The work of DJZ is supported by the NSFC Grant (12271334, 12671301) and the NSFC Bilateral Program (1241540016).

\appendix

\section{Multi-component generalization}\label{sec-3-4}

The reduction of Section \ref{sec-3-1} and the Cauchy matrix construction of
Section \ref{sec-4-1} extend to $G=\mathrm{GL}(M,\mathbb C)$ with no essential change. Theorem \ref{thm-CMA} is already formulated for $\mathrm{GL}(M,\mathbb C)$ and therefore
supplies Cauchy-matrix data for the corresponding matrix ASDYM solutions.
The additional reductions needed for the Heisenberg and WKI-type equations are stated below.
We give the NLS and GI branches and the Heisenberg/matrix-WKI-type branch explicitly. The full
matrix GDNLS family and the (2+1)-dimensional matrix Myrzakulov-I equation are not needed below,
so their explicit forms are omitted.

Let $M=M_1+M_2$ and replace $C$ in \eqref{dimensional-reduction} by
	\begin{align}\label{E-block}
		C=\frac{1}{2}\mathrm{diag}(I_{M_1},-I_{M_2})\in\mathfrak{gl}(M,\mathbb C),
	\end{align}
writing $J,K,P,Q$ in the corresponding $2\times2$ block form, e.g.
$K=(K_{ab})$ with $K_{ab}$ of size $M_a\times M_b$. Then
$[K,C]=\left(\begin{smallmatrix}0&-K_{12}\\ K_{21}&0\end{smallmatrix}\right)$
as in the $2\times2$ case. The derivation of \eqref{3.3a}--\eqref{3.3c} is unchanged except
that the blocks no longer commute. Setting again
$(r,q):=(K_{21},-K_{12})$, now of sizes $M_2\times M_1$ and $M_1\times M_2$, one gets
	\begin{align}
		K_{11}=-\partial_z^{-1}(qr),~~~~K_{22}=\partial_z^{-1}(rq)
	\end{align}
(note that $K_{11}=-K_{22}$ no longer holds when $M_1\neq M_2$) and the
\emph{multi-component (2+1)-d NLS system}
	\begin{subequations}\label{NLS-system-M}
		\begin{align}
			&r_w=r_{z\tilde z}-\big[\partial_z^{-1}(rq)_{\tilde z}\big]r
			-r\big[\partial_z^{-1}(qr)_{\tilde z}\big], \\
			&q_w=-q_{z\tilde z}+\big[\partial_z^{-1}(qr)_{\tilde z}\big]q
			+q\big[\partial_z^{-1}(rq)_{\tilde z}\big].
		\end{align}
	\end{subequations}
For $M_1=M_2=1$ this is \eqref{NLS-system}. Under $(x,y,t)=(\mathrm iz,\mathrm i\tilde z,\mathrm iw)$
and the conjugate reduction $q=r^\dagger$ it yields the matrix Zakharov equation
	\begin{align}
		\mathrm iq_t=q_{xy}+\big[\partial_x^{-1}(qq^\dagger)_y\big]q+q\big[\partial_x^{-1}(q^\dagger q)_y\big].
	\end{align}

Similarly, $p:=J_{21}J_{11}^{-1}$ (of size $M_2\times M_1$) satisfies the matrix Miura
transformation
	\begin{align}\label{Miura-M}
		p_z=-r+pqp,
	\end{align}
and substituting $r=-p_z+pqp$ into \eqref{NLS-system-M} gives, after using the identity
$\partial_z^{-1}(p_zq)_{\tilde z}+\partial_z^{-1}(pq_z)_{\tilde z}=(pq)_{\tilde z}$,
the \emph{multi-component (2+1)-d GI system}
	\begin{subequations}\label{GI-system-M}
		\begin{align}
			&p_w=p_{z\tilde z}-\big[\partial_z^{-1}(pq_z+pqpq)_{\tilde z}\big]p
			-p\big[\partial_z^{-1}(q_zp+qpqp)_{\tilde z}\big], \\
			&q_w=-q_{z\tilde z}-\big[\partial_z^{-1}(qp_z-qpqp)_{\tilde z}\big]q
			-q\big[\partial_z^{-1}(p_zq-pqpq)_{\tilde z}\big],
		\end{align}
	\end{subequations}
which reduces to \eqref{GI-system} when $M_1=M_2=1$. The gauge transformations
\eqref{gauge-trans-1}--\eqref{gauge-trans-2} also admit matrix counterparts, with $s=J_{11}$
now an $M_1\times M_1$ matrix obeying $(\partial_zs)s^{-1}=-qp$. We do not write out the
resulting multi-component GDNLS systems here.

For the Heisenberg branch one sets $\Sigma:=2E=\mathrm{diag}(I_{M_1},-I_{M_2})$ and
$S:=J^{-1}\Sigma J$, so that $S^2=I$, $\mathrm{Tr}(S)=M_1-M_2$, and
\eqref{Q-S}, \eqref{Q-w}, and \eqref{S-zt-evo} remain valid. Lemma \ref{lem-1} remains valid in
the form
	\begin{align}
		A_{\perp}=\frac{1}{4}[S,[S,A]]=\frac{1}{2}(A-SAS),~~~~
		A_{\parallel}=\frac{1}{2}(A+SAS),
	\end{align}
the expression $A_{\parallel}=\frac{1}{2}\mathrm{Tr}(AS)S$ being special to
$\mathfrak{sl}(2)$ (in general $A_\parallel$ need not be proportional to $S$). The argument of Theorem \ref{thm-1} then gives a matrix analogue of the Myrzakulov-I equation.
Only its
(1+1)-dimensional reduction is needed here. Setting $\partial_{\tilde z}=\partial_z$ and $A_\parallel=0$ in
$\partial_wQ=\frac{1}{4}[S,\partial_{\tilde z}S]+A_\parallel$, the compatibility condition
$\partial_z(\partial_wQ)=\partial_w(\partial_zQ)=-\frac{1}{2}\partial_wS$ gives
$\partial_wS=-\frac{1}{2}[S,\partial_z^2S]$, i.e.\ the matrix Heisenberg ferromagnet model
	\begin{align}\label{HF-matrix}
		\mathrm iS_t=\frac{1}{2}[S,S_{xx}],~~~~S=J^{-1}\Sigma J,~~~~S^2=I_M,~~~~
		\mathrm{Tr}(S)=M_1-M_2,
	\end{align}
which is \eqref{HFM-1p1} with $S$ now an $M\times M$ matrix. For $M_1=1$ one has
$S=-I_M+2P$ with $P$ a rank-one projector, and \eqref{HF-matrix} is the Heisenberg model on
$\mathbb{CP}^{M-1}$.

\begin{remark}\label{rem-no-matrix-WKI}
The WKI branch requires an additional restriction.
A reciprocal transformation introduces $\partial_X=-\omega^{-1}\partial_x$ for a single
scalar function $\omega$, so that the spectral problem \eqref{HF-Lax-a} becomes
$\partial_X\widehat\Psi=-\frac{\mathrm i\zeta}{2}\,\omega^{-1}S\,\widehat\Psi$.
To reach the WKI form \eqref{WKI-Lax} one needs $\omega^{-1}S$, possibly after a
$\zeta$-independent gauge transformation, to have diagonal blocks $I_{M_1}$ and $-I_{M_2}$.
Two conditions follow. First, taking traces and using $\mathrm{Tr}(S)=M_1-M_2$ gives
$(M_1-M_2)/\omega=M_1-M_2$, so that $\omega=1$ unless $M_1=M_2$. Thus a nonconstant
reciprocal factor, and hence a nontrivial WKI-type reduction of this two-block form, requires
$M_1=M_2$. Second, writing then
$\omega^{-1}S=\left(\begin{smallmatrix}I&-\mathrm i\hat q\\ -\mathrm i\hat r&-I\end{smallmatrix}\right)$,
the identity $(\omega^{-1}S)^2=\omega^{-2}I$ forces
	\begin{align}\label{QR-condition}
		\hat q\hat r=\hat r\hat q=\big(1-\omega^{-2}\big)I,~~~~\text{equivalently}~~~~
		\omega=\big(1-\hat q\hat r\big)^{-1/2},
	\end{align}
i.e.\ the product of the two potentials must be a multiple of the identity. Note that
\eqref{QR-condition} is the matrix analogue of \eqref{sqrt-identity}.

Condition \eqref{QR-condition} is the constraint imposed by Tsuchida and Wadati
\cite{Tsuchida-1999} on their multi-field WKI systems, which they realize by expanding $S$
in a Clifford algebra,
	\begin{align}\label{Clifford}
		S=\sum_{k=0}^{2m}s_ke_k,~~~~\{e_i,e_j\}=2\delta_{ij}I,~~~~M=2^m,~~~~
		\sum_{k=0}^{2m}s_k^2=1 .
	\end{align}
All the $e_k$ are traceless, so $M_1=M_2=2^{m-1}$. Choose a representation in which
\[
e_0=\mathrm{Diag}(I_{2^{m-1}},-I_{2^{m-1}})
\]
and $e_1,\dots,e_{2m}$ are block off-diagonal. Then $\omega^{-1}S$ has the WKI form with
$\omega=s_0$, and the scattering problem and \eqref{WKI-system-S} extend to matrix-valued
$\hat q$ and $\hat r$.

At the level of the equations of motion, the Clifford reduction has an obstruction. For $S$ of
the form \eqref{Clifford} one has
	\begin{align}
		S_x^2=\sum_{j,k}s_{j,x}s_{k,x}e_je_k
		=\frac{1}{2}\sum_{j,k}s_{j,x}s_{k,x}\{e_j,e_k\}
		=\Big(\sum_ks_{k,x}^2\Big)I,
	\end{align}
a scalar, so that the higher HF flow $S_t+S_{xxx}+\frac{3}{2}(S_x^2S)_x=0$ of
\cite{Tsuchida-1999} closes on the linear span of $\{e_k\}$. By contrast,
	\begin{align}
		[S,S_{xx}]=\sum_{j<k}\big(s_js_{k,xx}-s_ks_{j,xx}\big)[e_j,e_k]
	\end{align}
takes values in the \emph{bivector} part of the Clifford algebra, which coincides with the
vector part only when $2m+1=3$, i.e.\ $m=1$. For $m\geq2$ the two subspaces intersect
trivially. Hence, for $M\geq4$, the ansatz \eqref{Clifford} does not yield a nontrivial closed
reduction of the first flow \eqref{HFM-1p1}, and consequently does not produce a coupled-scalar
version of the WKI equation \eqref{Shimizu-Wadati}. This is consistent with the observation in the concluding remarks of \cite{Tsuchida-1999} that a simple
multi-component generalization of these first flows remains open. The standard Clifford ansatz therefore
closes for the higher flows but not for the first flow. Other reductions are not excluded.

The matrix Heisenberg model \eqref{HF-matrix} for arbitrary $M$ coincides with eq.~(2.5) of
\cite{Tsuchida-1999}, and \eqref{HF-Lax-1p1} is its Lax pair. Here both follow directly from the
ASDYM reduction. If one keeps $M_1=1$ instead of imposing
\eqref{QR-condition}, then $g$ is a row and $e$ a column, $ge$ is a scalar,
\eqref{sqrt-identity} survives, and $S=-I_M+2P$ with $P$ of rank one gives
$h=-I_{M-1}+\frac{eg}{1+f}$, so that \eqref{HF-Lax-a} becomes
	\begin{align}\label{vector-WKI-Lax}
		\partial_X\widehat\Phi=-\frac{\mathrm i\zeta}{2}
		\begin{pmatrix}
			1 & -\mathrm i\hat q \\[4pt]
			-\mathrm i\hat r & -\rho\, I_{M-1}-\dfrac{\hat r\hat q}{1+\rho}
		\end{pmatrix}\widehat\Phi,
		~~~~\rho:=\sqrt{1-\hat q\hat r}=\frac{1}{f}.
	\end{align}
For $M=2$ this is \eqref{WKI-Lax}, whereas for $M\geq3$ the lower-right block is not
proportional to the identity: the system lies outside the WKI class, in accordance with
\eqref{QR-condition}.
\end{remark}
\begin{proposition}\label{prop-matrix-WKI}
Let $S=J^{-1}\Sigma J$ satisfy the matrix Heisenberg model \eqref{HF-matrix}, and set
	\begin{align}\label{omega-H}
		\omega:=\frac{1}{M}\mathrm{Tr}(\Sigma S),~~~~
		\mathcal H:=\frac{\mathrm i}{2M}\mathrm{Tr}\big(\Sigma[S,S_x]\big).
	\end{align}
(The normalized trace is chosen so that $\omega=1$ at the vacuum $S=\Sigma$. For the
Clifford ansatz \eqref{Clifford} it gives $\omega=s_0$.)
Then $\partial_t\omega+\partial_x\mathcal H=0$, so that the reciprocal transformation
	\begin{align}\label{reciprocal-matrix}
		\mathrm dX=-\omega\,\mathrm dx+\mathcal H\,\mathrm dt,~~~~T=t
	\end{align}
is well defined wherever $\omega\neq0$, and in the variables $(X,T)$ the equation
\eqref{HF-matrix} becomes
	\begin{align}\label{matrix-WKI}
		\mathrm iS_T=\frac{1}{2}\omega^2[S,S_{XX}]+\frac{1}{2}\omega\omega_X[S,S_X]
		-\frac{\omega}{2M}\,\mathrm{Tr}\big(\Sigma[S,S_X]\big)\,S_X,
		~~~~\omega=\frac{1}{M}\mathrm{Tr}(\Sigma S),
	\end{align}
which is the compatibility condition of the Lax pair
	\begin{align}\label{matrix-WKI-Lax}
		\partial_X\Phi=-\frac{\mathrm i\zeta}{2\omega}S\,\Phi,~~~~
		\partial_T\Phi=\Big[\frac{\mathrm i\zeta^2}{2}S-\frac{\zeta\omega}{4}[S,S_X]
		+\frac{\mathrm i\zeta}{2}\,\frac{\mathcal H}{\omega}\,S\Big]\Phi .
	\end{align}
For $M=2$ one has $\Sigma=\sigma_3$ and $\omega=f$, and
\eqref{matrix-WKI}--\eqref{matrix-WKI-Lax} are equivalent to the WKI system \eqref{WKI-system}
and its Lax pair \eqref{WKI-Lax}--\eqref{WKI-Lax-time}, respectively.
\end{proposition}

\begin{proof}
Taking the trace of the matrix Heisenberg equation \eqref{HF-matrix} against $\Sigma$ gives
\begin{align*}
\omega_t
&=\frac{1}{M}\mathrm{Tr}(\Sigma S_t)
=-\frac{\mathrm i}{2M}\mathrm{Tr}\big(\Sigma[S,S_{xx}]\big).
\end{align*}
On the other hand,
\begin{align*}
\mathcal H_x
&=\frac{\mathrm i}{2M}\mathrm{Tr}\big(\Sigma[S,S_x]_x\big)
=\frac{\mathrm i}{2M}\mathrm{Tr}\big(\Sigma[S,S_{xx}]\big),
\end{align*}
because $[S_x,S_x]=0$. Hence $\omega_t+\mathcal H_x=0$, so the one-form in
\eqref{reciprocal-matrix} is closed. On any simply connected domain on which $\omega\neq0$,
it therefore defines a local change of variables with
\begin{align}\label{matrix-reciprocal-derivatives}
\partial_x=-\omega\partial_X,
\qquad
\partial_t=\partial_T+\mathcal H\partial_X.
\end{align}
Consequently
\begin{align*}
S_x=-\omega S_X,
\qquad
S_{xx}=\omega^2S_{XX}+\omega\omega_XS_X,
\qquad
S_t=S_T+\mathcal H S_X.
\end{align*}
Substitution into \eqref{HF-matrix} yields
\begin{align*}
\mathrm iS_T
=\frac{1}{2}\omega^2[S,S_{XX}]
+\frac{1}{2}\omega\omega_X[S,S_X]
-\mathrm i\mathcal H S_X.
\end{align*}
Moreover, by \eqref{omega-H} and $S_x=-\omega S_X$,
\begin{align*}
\mathcal H=-\frac{\mathrm i\omega}{2M}
\mathrm{Tr}\big(\Sigma[S,S_X]\big),
\end{align*}
which gives \eqref{matrix-WKI}.

It remains to transform the Lax pair. In the real coordinates used above, the
$(1+1)$-dimensional Heisenberg Lax pair \eqref{HF-Lax-1p1} takes the form
\begin{align*}
\partial_x\Phi=\frac{\mathrm i\zeta}{2}S\Phi,
\qquad
\partial_t\Phi=
\left[\frac{\mathrm i\zeta^2}{2}S+\frac{\zeta}{4}[S,S_x]\right]\Phi .
\end{align*}
Using \eqref{matrix-reciprocal-derivatives}, $S_x=-\omega S_X$, and the expression for
$\mathcal H$ above, the first equation becomes
\begin{align*}
\partial_X\Phi=-\frac{\mathrm i\zeta}{2\omega}S\Phi,
\end{align*}
and the second becomes
\begin{align*}
\partial_T\Phi
=\left[\frac{\mathrm i\zeta^2}{2}S
-\frac{\zeta\omega}{4}[S,S_X]
+\frac{\mathrm i\zeta}{2}\frac{\mathcal H}{\omega}S\right]\Phi.
\end{align*}
These are \eqref{matrix-WKI-Lax}. Since they are obtained from
\eqref{HF-Lax-1p1} by an invertible local change of independent variables, their compatibility
is equivalent to \eqref{matrix-WKI}.
\end{proof}

\medskip
\noindent\textbf{Explicit two- and four-field cases.}
We first record the low-dimensional cases of \eqref{matrix-WKI} in the usual WKI variables.  For $M=2$, write
\begin{align}
 S=\begin{pmatrix} f&g\\ e&-f\end{pmatrix},\qquad f^2+eg=1,
 \qquad \hat q=\mathrm i\frac{g}{f},\qquad \hat r=\mathrm i\frac{e}{f}.
\end{align}
Then $\omega=f=(1-\hat q\hat r)^{-1/2}$, and \eqref{matrix-WKI} is the two-field
WKI system already obtained in \eqref{WKI-system},
\begin{subequations}\label{two-field-WKI}
\begin{align}
 \hat q_T&=\pm\mathrm i\left(\frac{\hat q}{\sqrt{1-\hat q\hat r}}\right)_{XX},\\
 \hat r_T&=\mp\mathrm i\left(\frac{\hat r}{\sqrt{1-\hat q\hat r}}\right)_{XX}.
\end{align}
\end{subequations}
The Hermitian reduction $\hat r=-\hat q^*$ gives the scalar complex WKI equation
\eqref{WKI-final}.  We use the term \emph{two-field} here to distinguish the pair
$(\hat q,\hat r)$ from other notions of multi-component WKI equations in the literature.

As a $4\times4$ example, take $M=4$ with $M_1=M_2=2$ and, after a constant
permutation of the basis if necessary, put
\begin{align}\label{four-field-S}
 \Sigma=\mathrm{diag}(\sigma_3,\sigma_3),\qquad
 S=\mathrm{diag}(S^{(1)},S^{(2)}),\qquad
 S^{(a)}=\begin{pmatrix}f_a&g_a\\ e_a&-f_a\end{pmatrix},\quad a=1,2,
\end{align}
where
\begin{align}\label{four-field-constraint}
 f_a^2+e_ag_a=1,\qquad a=1,2.
\end{align}
Thus \eqref{four-field-S} contains four independent scalar fields.  Define
\begin{align}\label{four-field-omega-C}
 \omega=\frac{f_1+f_2}{2},\qquad
 \mathcal C:=\sum_{b=1}^2\bigl(g_be_{b,X}-g_{b,X}e_b\bigr).
\end{align}
Since $\mathrm{Tr}(\Sigma[S,S_X])=2\mathcal C$, the $4\times4$ equation
\eqref{matrix-WKI} is equivalent to the following closed system, for $a=1,2$:
\begin{subequations}\label{four-field-WKI}
\begin{align}
 \mathrm i f_{a,T}={}&\frac{\omega^2}{2}
 \bigl(g_ae_{a,XX}-g_{a,XX}e_a\bigr)
 +\frac{\omega\omega_X}{2}\bigl(g_ae_{a,X}-g_{a,X}e_a\bigr)
 -\frac{\omega}{4}\mathcal C f_{a,X},\label{four-field-WKI-f}\\
 \mathrm i g_{a,T}={}&\omega^2\bigl(f_ag_{a,XX}-g_af_{a,XX}\bigr)
 +\omega\omega_X\bigl(f_ag_{a,X}-g_af_{a,X}\bigr)
 -\frac{\omega}{4}\mathcal C g_{a,X},\label{four-field-WKI-g}\\
 \mathrm i e_{a,T}={}&\omega^2\bigl(e_af_{a,XX}-f_ae_{a,XX}\bigr)
 +\omega\omega_X\bigl(e_af_{a,X}-f_ae_{a,X}\bigr)
 -\frac{\omega}{4}\mathcal C e_{a,X}.\label{four-field-WKI-e}
\end{align}
\end{subequations}
On a domain where $f_1f_2\neq0$, one may introduce four WKI-type fields
\begin{align}\label{four-field-qr}
 \hat q_a:=\mathrm i\frac{g_a}{f_a},\qquad
 \hat r_a:=\mathrm i\frac{e_a}{f_a},\qquad
 f_a=(1-\hat q_a\hat r_a)^{-1/2},\qquad a=1,2.
\end{align}
Hence \eqref{four-field-WKI}, together with \eqref{four-field-qr}, is a reduction with the
four scalar fields $\hat q_1,\hat r_1,\hat q_2,\hat r_2$.  Under the Hermitian reduction
$e_a=g_a^*$, it is a system for the two complex fields $\hat q_1,\hat q_2$, equivalently four
real components.  The diagonal reduction $S^{(1)}=S^{(2)}$ gives $\omega=f_1=f_2$ and reduces
\eqref{four-field-WKI} to the ordinary two-field WKI system \eqref{two-field-WKI}.
This is a reducible block reduction and therefore does not
contradict the obstruction in Remark \ref{rem-no-matrix-WKI} or the open problem noted in
\cite{Tsuchida-1999} concerning a simple irreducible coupled-scalar generalization of the
first WKI flow.

\begin{remark}
Equation \eqref{matrix-WKI} is defined for arbitrary $M_1$ and $M_2$. No condition of the type
\eqref{QR-condition} is needed because that condition enters only when one insists on presenting the scattering
problem \eqref{matrix-WKI-Lax} in the two-block form \eqref{WKI-Lax}, with $\hat q$ and $\hat r$
as the sole potentials. Equation \eqref{matrix-WKI} is therefore not of the coupled-scalar shape
sought in \cite{Tsuchida-1999}. Remark \ref{rem-no-matrix-WKI} shows only that the standard
Clifford-algebra ansatz does not produce such a shape for the first flow. Equation \eqref{matrix-WKI} nevertheless gives a matrix-valued reciprocal form obtained directly
from the ASDYM reduction. Since the passage from \eqref{HF-matrix} to
\eqref{matrix-WKI} is only a change of independent variables, every Cauchy-matrix solution from
Section \ref{sec-4} that also satisfies the one-dimensional symmetry
$\partial_{\tilde z}=\partial_z$ and the reduction $A_\parallel=0$ produces a solution of
\eqref{matrix-WKI} through $S=J^{-1}\Sigma J$ and the quadrature
\eqref{reciprocal-matrix}, on any domain where $\omega\neq0$.
\end{remark}

With $C$ given by \eqref{E-block}, the Lax pair \eqref{NLS-Lax} extends without change, \eqref{NLS-Lax-a} becoming the multi-component Zakharov--Shabat spectral problem, and so
does the Heisenberg pair \eqref{HF-Lax} once $\kappa S$ is replaced by $A_\parallel$. The
conjugations \eqref{DNLS-gauge} of the DNLS family, on the other hand, involve the matrix
$s=J_{11}$ and require the care already mentioned above. In Section \ref{sec-4}, the constraint \eqref{4.10} makes
sense for \eqref{E-block} without modification, the plane wave factors \eqref{phi-psi} are
replaced by $\phi_{is}=\exp\big(\pm\frac{1}{2}(\xi_iz+\tilde w)\big)\varrho_{is}(\xi_iw+\tilde z)$
with the sign determined by the block to which the index $s$ belongs, and Theorem
\ref{thm-tau-Lagrangian} holds unchanged: its proof uses only $\partial_z\Omega=\eta C\theta$,
$[[K,C],C]=\left(\begin{smallmatrix}0&K_{12}\\ K_{21}&0\end{smallmatrix}\right)$ and
	\begin{align}
		\partial_z\mathrm{tr}(K_{11})=\mathrm{tr}(K_{12}K_{21})
		=\partial_z\mathrm{tr}(KC)=\partial_z^2(\ln\tau),
	\end{align}
so that $\mathcal L_{[\mathrm{ASD}]}=4\partial_{\tilde z}^2\partial_z^2(\ln\tau)$ for
$\mathrm{GL}(M,\mathbb C)$ as well.

\section{Proof of Theorem~\ref{thm-CMA}}\label{app-1}

We first prove the difference-to-product formula \eqref{diff-to-prod}. Using the definition \eqref{Sij} and the Sylvester equation \eqref{Cauchy-matrix-structure-SE}, we compute
\begin{align*}
	S^{(i+1,j)}-S^{(i,j+1)}
	&=
	\theta\Lambda^j\Omega^{-1}\Xi^{i+1}\eta
	-
	\theta\Lambda^{j+1}\Omega^{-1}\Xi^i\eta        \\
	&=
	\theta\Lambda^j
	\left(\Omega^{-1}\Xi-\Lambda\Omega^{-1}\right)
	\Xi^i\eta                                      \\
	&=
	\left(\theta\Lambda^j\Omega^{-1}\eta\right)
	\left(\theta\Omega^{-1}\Xi^i\eta\right) =
	S^{(0,j)}S^{(i,0)}.
\end{align*}
Therefore, the master functions satisfy
\[
S^{(i+1,j)}-S^{(i,j+1)}
=
S^{(0,j)}S^{(i,0)}.
\]

To prove the evolution relations \eqref{Sij-evo}, we first record differential identities for $\Omega$.
\begin{lemma}
	The matrix $\Omega$ in \eqref{Cauchy-matrix-structure} satisfies
	\begin{subequations}
		\begin{align}
			\label{omega-evo-1}&\partial_{z}\Omega=(\partial_{\tilde w}\eta)\theta+(\partial_{\tilde w}\Omega)\Lambda=\Xi(\partial_{\tilde w}\Omega)-\eta(\partial_{\tilde w}\theta), \\
			\label{omega-evo-2}&\partial_{w}\Omega=(\partial_{\tilde z}\eta)\theta+(\partial_{\tilde z}\Omega)\Lambda=\Xi(\partial_{\tilde z}\Omega)-\eta(\partial_{\tilde z}\theta).
		\end{align}
	\end{subequations}
\end{lemma}
\begin{proof}
	Taking the $z$-derivative of \eqref{Cauchy-matrix-structure-SE} gives
	\[
	\Xi(\partial_z\Omega)
	-
	(\partial_z\Omega)\Lambda
	=
	(\partial_z\eta)\theta
	+
	\eta(\partial_z\theta).
	\]
	Substituting \eqref{4.1b} into it yields
	\begin{align*}
		\Xi(\partial_z\Omega)
		-
		(\partial_z\Omega)\Lambda
		&=
		\Xi(\partial_{\tilde w}\eta)\theta
		+
		\eta(\partial_{\tilde w}\theta)\Lambda \\
		&=
		\Xi((\partial_{\tilde w}\eta)\theta+(\partial_{\tilde w}\Omega)\Lambda)
		-
		(\Xi(\partial_{\tilde w}\Omega)-\eta(\partial_{\tilde w}\theta))\Lambda.
	\end{align*}
	The $\tilde w$-derivative of \eqref{Cauchy-matrix-structure-SE} is equivalent to
	\[
	(\partial_{\tilde w}\eta)\theta+(\partial_{\tilde w}\Omega)\Lambda
	=
	\Xi(\partial_{\tilde w}\Omega)-\eta(\partial_{\tilde w}\theta).
	\]
	Thus the two matrices appearing on the right-hand side of the preceding Sylvester equation
	coincide. The disjoint-spectrum assumption then implies, by uniqueness of the homogeneous
	Sylvester problem, the two expressions in \eqref{omega-evo-1}. The proof of
	\eqref{omega-evo-2} is identical after replacing $(z,\tilde w)$ by $(w,\tilde z)$.

\end{proof}

Next, differentiating $S^{(i-1,j)}$ with respect to $z$ gives
\begin{align*}
	\partial_zS^{(i-1,j)}
	=
	(\partial_z\theta)\Lambda^j\Omega^{-1}\Xi^{i-1}\eta
	-
	\theta\Lambda^j\Omega^{-1}
	(\partial_z\Omega)
	\Omega^{-1}\Xi^{i-1}\eta
	+
	\theta\Lambda^j\Omega^{-1}\Xi^{i-1}
	(\partial_z\eta).
\end{align*}
Substituting \(\partial_z\eta=\Xi(\partial_{\tilde w}\eta)\) and \(\partial_z\theta=(\partial_{\tilde w}\theta)\Lambda\) gives
\begin{align*}
	\partial_zS^{(i-1,j)}
	=
	(\partial_{\tilde w}\theta)\Lambda^{j+1}
	\Omega^{-1}\Xi^{i-1}\eta
	-
	\theta\Lambda^j\Omega^{-1}
	(\partial_z\Omega)
	\Omega^{-1}\Xi^{i-1}\eta
	+
	\theta\Lambda^j\Omega^{-1}\Xi^i
	(\partial_{\tilde w}\eta).
\end{align*}
The $\tilde w$-derivative of $S^{(i,j)}$ is
\begin{align*}
	\partial_{\tilde w}S^{(i,j)}
	=
	(\partial_{\tilde w}\theta)\Lambda^j
	\Omega^{-1}\Xi^i\eta
	-
	\theta\Lambda^j\Omega^{-1}
	(\partial_{\tilde w}\Omega)
	\Omega^{-1}\Xi^i\eta
	+
	\theta\Lambda^j\Omega^{-1}\Xi^i
	(\partial_{\tilde w}\eta).
\end{align*}
Thus
\begin{align*}
	\partial_zS^{(i-1,j)}
	-
	\partial_{\tilde w}S^{(i,j)}
	&=
	(\partial_{\tilde w}\theta)\Lambda^j
	\left(\Lambda\Omega^{-1}-\Omega^{-1}\Xi\right)
	\Xi^{i-1}\eta                                             \\
	&\quad
	-
	\theta\Lambda^j\Omega^{-1}
	(\partial_z\Omega)
	\Omega^{-1}\Xi^{i-1}\eta                                  \\
	&\quad
	+
	\theta\Lambda^j\Omega^{-1}
	(\partial_{\tilde w}\Omega)
	\Omega^{-1}\Xi^i\eta.
\end{align*}
Define
\begin{align*}
	&A:=	(\partial_{\tilde w}\theta)\Lambda^j
	\left(\Lambda\Omega^{-1}-\Omega^{-1}\Xi\right)
	\Xi^{i-1}\eta, \\
	&B:=\theta\Lambda^j\Omega^{-1}
	(\partial_z\Omega)
	\Omega^{-1}\Xi^{i-1}\eta, \\
	&C:=\theta\Lambda^j\Omega^{-1}
	(\partial_{\tilde w}\Omega)
	\Omega^{-1}\Xi^i\eta.
\end{align*}
Substituting $\Lambda\Omega^{-1}-\Omega^{-1}\Xi=-\Omega^{-1}\eta\theta\Omega^{-1}$ into the definition of $A$ yields
\begin{align*}
	A=-\left((\partial_{\tilde w}\theta)\Lambda^j
	\Omega^{-1}\eta\right)\left(\theta\Omega^{-1}
	\Xi^{i-1}\eta\right).
\end{align*}
Likewise, $C-B$ is expressed via \eqref{omega-evo-1} as
\begin{align*}
	C-B=\,
	\theta\Lambda^j\Omega^{-1}
	(\partial_{\tilde w}\Omega)
	\Omega^{-1}\eta\theta\Omega^{-1}
	\Xi^{i-1}\eta
	-
	\theta\Lambda^j\Omega^{-1}
	(\partial_{\tilde w}\eta)\theta
	\Omega^{-1}\Xi^{i-1}\eta.
\end{align*}
Hence
\begin{align*}
	A-B+C=-((\partial_{\tilde w}\theta)\Lambda^j
	\Omega^{-1}\eta-\theta\Lambda^j\Omega^{-1}
	(\partial_{\tilde w}\Omega)
	\Omega^{-1}\eta+\theta\Lambda^j\Omega^{-1}
	(\partial_{\tilde w}\eta))\theta
	\Omega^{-1}\Xi^{i-1}\eta.
\end{align*}
Using the expressions for $\partial_{\tilde w}S^{(0,j)}$ and $S^{(i-1,0)}$, we obtain
\begin{align*}
	\partial_zS^{(i-1,j)}
	-
	\partial_{\tilde w}S^{(i,j)}
	=
	A-B\mathbin{+}C
	=
	-\partial_{\tilde w}S^{(0,j)}S^{(i-1,0)}.
\end{align*}
This proves \eqref{Sij-evo-2}. The proof of \eqref{Sij-evo-1} is analogous.

Finally, consider the determinant formula for $J$:
\begin{align*}
	\det(J)
	&=
	\det
	\left(I-\theta\Omega^{-1}\Xi^{-1}\eta\right) =
	\det
	\left(I-\Omega^{-1}\Xi^{-1}\eta\theta\right)                                                   \\
	&=
	\det
	\left(I-\Omega^{-1}\Xi^{-1}
	\left(\Xi\Omega-\Omega\Lambda\right)\right)                                                   \\
	&=
	\det
	\left(I-\Omega^{-1}\Omega
	+
	\Omega^{-1}\Xi^{-1}\Omega\Lambda\right)                                                   \\
	&=
	\det
	\left(\Omega^{-1}\Xi^{-1}\Omega\Lambda\right)                                                   \\
	&=
	\frac{\det(\Lambda)}{\det(\Xi)},
\end{align*}
and the trace formula for $K$:
\begin{align*}
	\tr(K)
	&=
	\tr
	\left(\theta\Omega^{-1}\eta\right) =
	\tr
	\left(\Omega^{-1}\eta\theta\right)                                                   \\
	&=
	\tr
	\left(\Omega^{-1}
	\left(\Xi\Omega-\Omega\Lambda\right)\right)                                                   \\
	&=
	\tr
	\left(\Omega^{-1}\Xi\Omega\right)
	-
	\tr
	\left(\Lambda\right)                                                   \\
	&=
	\tr
	\left(\Xi\right)
	-
	\tr
	\left(\Lambda\right).
\end{align*}
The formulas for $\det(P)$ and $\tr(Q)$ follow similarly.

\bigskip
\begingroup
\small
\emergencystretch=2em

\endgroup

\end{document}